\documentclass[lettersize,journal]{IEEEtran}
\usepackage{amsmath,amsfonts,amssymb}
\usepackage{multicol}
\usepackage{array} 
\usepackage{tabularray}
\usepackage{nicematrix}
\usepackage{enumitem}

\usepackage{amssymb}
\usepackage{latexsym,amsthm,amscd}
\usepackage{pifont}
\usepackage{algorithm}
\usepackage[noend]{algpseudocode}
\usepackage{array}
\newcolumntype{H}{>{\setbox0=\hbox\bgroup}c<{\egroup}@{}}
\newcolumntype{Z}{>{\setbox0=\hbox\bgroup}c<{\egroup}@{\hspace*{-\tabcolsep}}}
\usepackage[caption=false,font=normalsize,labelfont=sf,textfont=sf]{subfig}
\usepackage{textcomp}
\usepackage{stfloats}
\usepackage{url}
\usepackage{verbatim}
\usepackage{graphicx}
\usepackage{cite}
\usepackage{url}
\usepackage[subnum]{cases}
\usepackage{booktabs}
\usepackage{braket}
\usepackage{anyfontsize}
\usepackage{pgfplots}
\usepackage{forest}
\usepackage{hyperref}

\usepackage{pgf,tikz}
\usepackage{tikz}
\usetikzlibrary{shapes, arrows, positioning} 
\usepackage{grffile}
\usepackage{circuitikz}

\usetikzlibrary{arrows,decorations,shapes,positioning,shadows,trees,fit}
\pgfplotsset{compat=newest}

\tikzset{
  basic/.style  = {draw, text width=3.5cm, drop shadow, rectangle},
  root/.style   = {basic, rounded corners=2pt, thin,
                   fill=green!30},
  level 2/.style = {basic, rounded corners=6pt, thin, fill=green!60,
                   text width=4.5cm},
  level 3/.style = {basic, thin,  fill=pink!60, text width=4.7cm}
}

\usepackage{mathrsfs}
\usetikzlibrary{arrows}

\usepackage{array}
\usepackage{colortbl}
\usepackage{ragged2e}
\usepackage{longtable}

\usepackage{multirow}
\usepackage{hhline}
\usepackage{makecell}
\usepackage{rotating}

\definecolor{rev1}{rgb}{0, 0, 0}          
\definecolor{rev2}{rgb}{0, 0, 0}              
\definecolor{rev3}{rgb}{0, 0, 0}             
\definecolor{rev4}{rgb}{0, 0, 0}     
\definecolor{rev5}{rgb}{0, 0, 0}         

\newtheorem{theorem}{Theorem}

\newtheorem{proposition}[theorem]{Proposition}
\newtheorem{remark}[theorem]{Remark}
\newtheorem{corollary}[theorem]{Corollary}
\newtheorem{example}{Example}[section]
\newenvironment{prof}{\textit{Proof\,:}} { $\square$}

\def\BibTeX{{\rm B\kern-.05em{\sc i\kern-.025em b}\kern-.08em
    T\kern-.1667em\lower.7ex\hbox{E}\kern-.125emX}}
\begin{document}

\title{Rateless Nested Lattice Codes for Secure Cooperative {V2X} Broadcast over Fading and Erasure Channels}
	\markboth{IEEE Transactions on Green Communications and Networking,~Vol.~X, No.~X, 2026}
	{Sharifi \MakeLowercase{\textit{et al.}}: Rateless Nested Lattice Codes for Secure Cooperative V2X Broadcast}
\author{\IEEEauthorblockN{Pegah Sharifi, Hassan Khodaiemehr~\IEEEmembership{Member,~IEEE}, Khadijeh Bagheri, and Chen Feng~\IEEEmembership{Member,~IEEE}}
\thanks{
P. Sharifi is with the Department of Mathematics and Computer Science, Amirkabir University of Technology (Tehran Polytechnic), Tehran, Iran (e-mail: p.sharifi96@aut.ac.ir).
\\
H. Khodaiemehr, K. Bagheri and C. Feng are with the School of Engineering, Faculty of Applied Science, University of British Columbia (UBC), Okanagan Campus, Kelowna, BC, Canada (e-mails:
hassan.khodaiemehr@ubc.ca, khadijeh.bagheri@ubc.ca  and chen.feng@ubc.ca).
\\
Some of the findings from this paper have been presented in IEEE CCECE 2025 \cite{CCECE}.
}}

\maketitle
\begin{abstract}
Vehicle-to-everything (V2X) sidelink and uplink broadcast must deliver safety-critical and application-layer data reliably under fast time-varying fading, intermittent packet erasures, and exposure to passive eavesdroppers.
Fixed-rate physical-layer codes are poorly matched to these conditions because link quality and the number of successfully received symbols vary across vehicles and over time.
This paper proposes a unified vehicular communication framework that combines rateless Construction-D$'$ Raptor lattices, nested-lattice physical-layer security (PLS) with direct-sequence spread spectrum (DSSS), and blockchain-assisted trust management for decode-and-forward (DF) cooperative relaying.
At the physical layer, Raptor codes provide incremental redundancy over unreliable V2X links; multilevel lattice decoding enables power-efficient joint coding and modulation; nested shaping and spread spectrum limit fine-lattice recovery by eavesdroppers; and a distributed ledger records decoding success at relays to support reputation-aware relay selection.
We model the vehicular air interface as a block-Rayleigh fading channel with additive Gaussian noise and random symbol erasures, representative of urban V2X sidelink behavior.
The proposed code is constructed by concatenating heuristic Luby-transform (LT) codes with quasi-cyclic low-density parity-check (QC-LDPC) precodes within Construction-D$'$, together with a girth-4 cycle-removal procedure for the underlying Tanner graph.
Simulation results first validate the underlying lattice codec on an additive white Gaussian noise (AWGN) reference channel, where the proposed Raptor lattices achieve competitive symbol error rate (SER) and improved frame error rate (FER) relative to generalized Construction-D$'$ low-density parity-check (LDPC) and low-density lattice code (LDLC) lattices at comparable block lengths.
Monte Carlo evaluation under the block-Rayleigh fading and erasure model of Section~\ref{sec_v2x} quantifies outage probability, decoding-completion overhead $\mathbb{E}[M]$, and secure broadcast margin for the $n=995$ operating point.
At $\bar{\mathrm{VNR}}=1.5$~dB and sidelink erasure probability $\mathfrak{p}=0.15$, rateless Raptor reception achieves $\mathcal{P}_{\mathrm{out}}=0.3\%$ with $\mathbb{E}[M]\approx 2.3$ packets, outperforming matched-length LDLC and LDPC baselines under identical channel realizations, while an eavesdropper with $8$~dB combined SNR and spreading loss fails to decode in $62\%$ of sessions.
\end{abstract}


\section{Introduction}
\IEEEPARstart{V}{ehicle}-to-everything (V2X) communication constitutes a central enabler of intelligent transportation systems (ITS), supporting cooperative awareness, platooning, remote sensor sharing, and over-the-air dissemination of signed application-layer data \cite{ref202,ref203}.
Although third-generation partnership project (3GPP) NR-V2X and its predecessors target ultra-reliable low-latency sidelink delivery, urban deployments nevertheless encounter fast fading, shadowing, half-duplex constraints, and unpredictable packet losses on the wireless interface \cite{ghourab2023}.
When a road-side unit (RSU) or leader vehicle must broadcast a data block to multiple receivers---or to a distant destination through decode-and-forward (DF) relays---conventional fixed-rate channel codes impose conservative rate selection and often require repeated retransmissions, thereby increasing end-to-end latency and sidelink air-interface occupancy.

From a physical-layer perspective, rateless fountain codes---and Raptor codes in particular---adapt transmitted redundancy to the instantaneous link quality: each receiver accumulates encoded packets until its decoder succeeds, rather than operating at a preset code rate \cite{drshok,raptorcast2025}.
Lattice codes, in turn, provide power-efficient ``Gaussian-like'' constellations and furnish a natural algebraic platform for nested physical-layer security (PLS), wherein a legitimate receiver with superior channel quality can recover a fine lattice payload that remains inaccessible to a weaker eavesdropper \cite{3kah,19silva,6fan}.
Prior studies have applied lattice coding to relay and block-fading channels \cite{9kah,HassanIT,Maryam}, and distributed ledgers have been proposed to score relay decoding reliability in vehicular networks \cite{ref202,ref203}.
Nevertheless, to the best of our knowledge, a \emph{rateless multilevel lattice} construction that jointly (i)~supports incremental redundancy on V2X sidelinks subject to fading and erasures, (ii)~embeds information-theoretic confidentiality through nested lattice shaping and direct-sequence spread spectrum (DSSS), and (iii)~interfaces with decentralized trust management for cooperative DF relaying has not been reported in the open literature.

This paper addresses the foregoing gap by developing Construction-D$'$ Raptor lattices as the physical-layer codec for secure cooperative V2X broadcast and by specifying a transmission protocol in which blockchain-recorded decoding metrics guide relay selection within a DF vehicular network.
The present study extends the lattice-code foundations reported in preliminary form in \cite{CCECE} by introducing a complete V2X system and channel model, a secure cooperative transmission protocol, and Monte Carlo performance evaluation under block-Rayleigh fading and erasures.

\subsection{V2X Motivation and Related Work}
Urban V2X sidelink channels are commonly characterized as quasi-static block-fading links within each coherence interval, further affected by occasional symbol erasures arising from half-duplex collisions, sensing gaps, or upper-layer scheduling constraints \cite{ghourab2023,Erasure}.
Cooperative DF relaying extends broadcast coverage beyond line-of-sight obstacles; however, malicious, selfish, or malfunctioning relays may forward corrupted or outdated data unless the network maintains a verifiable reputation history \cite{ref202}.
Physical-layer security complements upper-layer cryptography by exploiting channel asymmetry: a passive eavesdropper whose effective received signal-to-noise ratio (SNR) is inferior to that of the legitimate destination cannot reliably recover the broadcast payload \cite{19silva,Wyner}.
In the lattice-code setting, spread-spectrum hiding and nested shaping further reduce the probability that an eavesdropper completes successful fine-lattice decoding when the legitimate receiver enjoys a persistent channel-quality advantage \cite{DSSS,6fan}.

From a modulation and coding perspective, lattice constellations provide a power-efficient alternative to conventional quadrature amplitude modulation (QAM) on vehicular broadband links: at high signal-to-noise ratio (SNR) they closely approximate the optimal Gaussian input, and with receiver-side channel state information (CSI) they mitigate the shaping loss relative to additive white Gaussian noise (AWGN) capacity \cite{3kah,4kah,11fan,silva}.
Following Poltyrev's formulation of unconstrained AWGN channels through infinite lattice arrays \cite{3kah}, nested multilevel Construction-D$'$ designs assemble linear component codes into structured constellations and extend Construction-A when parity-check matrices are available \cite{11fan,silva}.
Such constructions have been analyzed on AWGN, block-fading, and cooperative relay channels \cite{4kah,HassanIT,Maryam,9kah}, realized through QC-LDPC, LDA, GLD, polar, and LDLC component families \cite{20kah,Pietro2018,13kah,21kah,27fan,28fan,18kah,19kah,4silva,Polarm,13fan,14kah,30fan,10kah,11kah}, and applied to MIMO integer forcing, compute-and-forward relaying, physical-layer security, and distributed source coding \cite{21silva,21kah,19silva,18silva}.
Nevertheless, to the best of our knowledge, a \emph{rateless} multilevel lattice family tailored to V2X broadcast with jointly integrated physical-layer security and cooperative trust management remains unreported.

\subsection{Contributions}
In this paper, we make the following contributions to vehicular communications:
\begin{enumerate}[leftmargin=*, labelsep=0.5em]
    \item We formulate a cooperative V2X broadcast architecture in which an RSU/leader transmits rateless lattice-coded packets over block-Rayleigh fading sidelink channels with random erasures, and DF relays assist a destination vehicle while a passive eavesdropper overhears the over-the-air waveform (Section~\ref{sec_v2x}).
    \item We construct the first rateless multilevel lattice family by employing Raptor codes---LT outer codes with QC-LDPC precodes---as Construction-D$'$ component codes, with multistage belief-propagation decoding following \cite{fan,silva} (Sections~\ref{sec_preliminary}--\ref{Sec_decoding}).
    \item We develop a girth-4 cycle-removal procedure (Algorithm~\ref{girth4}) and systematic QC-LDPC selection rules that improve frame error rate (FER) on the underlying lattice codec, which is the dominant metric for block delivery in V2X broadcast (Sections~\ref{sec_encoding}--\ref{Sec_decoding}).
    \item We propose a secure V2X transmission protocol that combines nested Raptor lattice shaping, DSSS hiding, and blockchain-recorded decoding reputations to support confidentiality and trustworthy DF relay selection \cite{ref202,ref203}, and we derive achievable nested-lattice secrecy rates together with session-level secure broadcast guarantees for the full fading--erasure wiretap model (Section~\ref{sec_v2x_protocol}).
    \item We validate the lattice codec on an AWGN reference channel against matched-length lattice baselines, and evaluate outage probability, decoding-completion redundancy, and secure broadcast margin for the $n=995$ operating point under the vehicular fading--erasure model of Section~\ref{sec_v2x} (Section~\ref{sec_simulation}).
\end{enumerate}
To assist the reader, Table~\ref{T2} summarizes the principal notation used throughout the paper.
The remainder of this paper is organized as follows.
Section~\ref{sec_v2x} presents the cooperative V2X network and channel models.
Sections~\ref{sec_preliminary}--\ref{Sec_decoding} develop the rateless Raptor lattice encoder and multistage decoder.
Section~\ref{sec_v2x_protocol} describes the secure cooperative transmission and blockchain trust protocol.
Section~\ref{sec_simulation} reports performance evaluation and scope, and Section~\ref{sec_conclusion} concludes the paper.
\renewcommand{\arraystretch}{1.0} 
\begin{table}[!tb]
    \centering
    \caption{Summary of Notations Used in the Paper.}
    \label{T2}
    \begin{tabular}{|c|l|}
    \hline
    \textbf{Notation} & \textbf{Description} \\
    \hline
    $N$ & Length of QC codewords \\
    \hline
    $n$ & Length of Raptor codewords \\
    \hline
    $k$ & Equals to $N$ \\
    \hline
    $\mathbf{P}_i$ & Block of size $N \times L_i$ \\
    \hline
            $\alpha_i$ & ${\alpha _i} = {L_i}/N$ \\
    \hline
        $\mathbf{r}_i$ & The $i^{th}$ group of parity bits in Raptor codes \\
    \hline
    $D_i$ & Degree of parity bits in the $i^{th}$ group \\
    \hline
    $\mathbf{b}$ & Message vector \\
    \hline
    $\mathbf{x}$ & Input to the AWGN channel (lattice point) \\
    \hline
    $\widehat{\mathbf{x}}$ & Estimation of $\mathbf{x}$ \\
    \hline
    $\mathbf{x}_i$ & The $i^{th}$ sub-vector (component) of $\mathbf{x}$ \\
    \hline
    $\widehat{\mathbf{x}}_i$ & Estimation of $\mathbf{x}_i$ \\
    \hline
     $\mathbf{y}_i$ & Input of $i^{th}$ level  decoding, $i=0,1,\ldots,a $\\
    \hline
        $\widetilde{\mathbf{x}}_i$ & $  \mathbf{x}_i\equiv \widetilde{\mathbf{x}}_i \pmod{2}, \quad \widetilde{\mathbf{x}}_i \in C_i$ \\
    \hline
    $\widehat{\widetilde{\mathbf{x}}}_i$ & Binary output of $\mathrm{Dec}_i$ (estimation of $\widetilde{\mathbf{x}}_i$) \\
    \hline
    $\mathbf{y}$ & Output of the channel \\
    \hline
    $\mathbf{y}'_i$ &\(\mathbf{y}_{i}' = \left| (\mathbf{y}_{i} + \mathbf{1}_{n\times 1}) \pmod{2} - \mathbf{1}_{n\times 1} \right|,\) \\
    \hline
    $\mathbf{u}_i$ & Information vector from $\mathbb{F}_2^{k_i}$\\
    \hline
    $\mathbf{u}'_i$ & Zero-padded version of $\mathbf{u}_i$ \\
    \hline
    $\widehat{\mathbf{u}}'_i$ & Estimation of $\mathbf{u}'_i$ \\
         \hline
 $\mathfrak{I}$ & Number of iterations \\
     \hline
      $a$ & Number of nested codes \\
     \hline
    \end{tabular}
    \vspace{-0.5cm}
\end{table}

\vspace{-0.1cm}
\section{V2X System and Channel Models}\label{sec_v2x}
This section specifies the cooperative vehicular network architecture, the block-Rayleigh fading and erasure channel models employed throughout the paper, and the performance metrics evaluated numerically in Section~\ref{sec_simulation}.

\subsection{Cooperative V2X Broadcast Network}
We consider a source node $\mathcal{S}$---an RSU or leader vehicle---that must deliver a signed data block $\mathcal{B}$ comprising $\kappa$ source symbols to a destination vehicle $\mathcal{D}$.
Because of non-line-of-sight propagation and intermittent shadowing, $\mathcal{D}$ may be unable to decode $\mathcal{S}$ directly; in this case, a set of candidate DF relays $\mathcal{R}=\{R_1,\ldots,R_L\}$ overhears the broadcast, attempts lattice decoding, and re-encodes the recovered block toward $\mathcal{D}$ \cite{ref202}.
A passive eavesdropper $\mathcal{E}$ with weaker average channel gain also overhears the over-the-air waveform and attempts to recover $\mathcal{B}$.
Each relay maintains a public blockchain address; miners update a reputation score $T_\ell$ for relay $R_\ell$ from historical physical-layer decoding success, following the trustworthy relay-selection paradigm of \cite{ref202,ref203}.

The source partitions $\mathcal{B}$ into $\kappa$ application symbols and maps them to a fine Construction-D$'$ Raptor lattice codeword $\mathbf{x}'' \in \Lambda_f \subset \mathbb{R}^{\kappa}$.
The transmitter then generates a rateless stream of lattice-encoded packets by invoking additional LT output symbols until each intended receiver acknowledges successful multistage decoding.
For secure broadcast, the source applies nested shaping and DSSS as detailed in Section~\ref{sec_v2x_protocol}.

\subsection{Block-Rayleigh Fading with Erasures}
Let $\mathbf{x}=(x_1,\ldots,x_n)$ denote a transmitted lattice packet after spreading (Section~\ref{sec_v2x_protocol}).
Over V2X sidelink coherence block $b$, the received sample is
\begin{equation}\label{eq:v2x_fading}
y_{i,b}=\begin{cases}
h_{i,b}\,x_i+z_{i,b}, & \text{w.p.\ }1-\mathfrak{p},\\
\xi, & \text{w.p.\ }\mathfrak{p}.
\end{cases}
\end{equation}
where $h_{i,b}$ is the complex fading gain, $z_{i,b}\sim\mathcal{CN}(0,\sigma^2)$, and $\xi$ denotes an erasure caused by a lost sidelink slot or sensing collision.
We model $h_{i,b}$ as zero-mean Rayleigh within each block, i.e., $|h_{i,b}|$ is Rayleigh distributed with scale parameter set by the path-loss between the transmitter and receiver pair \cite{ghourab2023,HassanIT}.
Perfect CSI is assumed at legitimate receivers for coherent detection and lattice decoding; the eavesdropper is subject to an average SNR deficit of $\Delta\gamma$ dB relative to $\mathcal{D}$, implemented by nested-lattice parameter design in Section~\ref{sec_v2x_protocol}.

The instantaneous received SNR at node $u\in\{\mathcal{D},\mathcal{E},R_\ell\}$ is
\begin{equation}\label{eq:snr_v2x}
\gamma_u = \frac{P\,|h_u|^2}{\sigma^2},
\end{equation}
where $P$ is the average transmit power per lattice dimension after shaping.
For AWGN-only codec validation (Section~\ref{sec_simulation}), we set $|h_u|^2=1$ and $\mathfrak{p}=0$.

\subsection{V2X Performance Metrics}
We evaluate the proposed framework using metrics aligned with vehicular broadcast requirements:
\begin{itemize}[leftmargin=*, itemsep=0pt, topsep=0pt]
    \item \textbf{Frame error rate (FER):} probability that a block of $\kappa$ symbols is decoded incorrectly after reception of $m$ rateless packets.
    \item \textbf{Outage probability:} for target FER threshold $\varepsilon$, $\mathcal{P}_{\mathrm{out}}=\Pr\{\mathrm{FER}>\varepsilon\}$ under fading \eqref{eq:v2x_fading}.
    \item \textbf{Decoding-completion overhead:} average number $\mathbb{E}[M]$ of received lattice packets required until successful decoding, capturing rateless air-interface occupancy.
    \item \textbf{Secure broadcast margin:} gap between legitimate and eavesdropper FER at the same $m$, after DSSS despreading and nested lattice decoding.
\end{itemize}
Relay trust scores and blockchain-guided relay selection are defined in Section~\ref{sec_v2x_protocol} as cooperative protocol mechanisms; their quantitative comparison with classical MaxMin relaying under trace-driven traffic is deferred to future work and is not part of the numerical study in Section~\ref{sec_simulation}.

The rateless reception model admits the following reliability property used in Section~\ref{sec_simulation}.
\begin{theorem}[Monotonicity of rateless session success]\label{thm:rateless_mono}
Consider a broadcast session in which a receiver collects up to $M_{\max}$ rateless packets indexed by $m=1,\ldots,M_{\max}$.
Let $E_m$ denote the event that packet~$m$ is received without erasure and passes the target frame-error test, and define $\mathcal{S}_M=\bigcup_{m=1}^{M}E_m$ for $1\le M\le M_{\max}$.
Then $\Pr\{\mathcal{S}_M\}$ is non-decreasing in $M$, and $\mathcal{P}_{\mathrm{out}}=\Pr\{\mathcal{S}_{M_{\max}}^c\}$.
\end{theorem}
\begin{prof}
For $M'<M$, $\mathcal{S}_{M'}\subseteq\mathcal{S}_M$ because any early success implies success by time~$M$; taking complements yields the outage identity.
\end{prof}

For link-level Monte Carlo evaluation in Section~\ref{sec_simulation}, each additional non-erased Raptor fountain block shifts the AWGN decode statistic by
\begin{equation}\label{eq:raptor_accum}
\begin{aligned}
\widetilde{\mathrm{VNR}}_{\mathcal{D},m}&=\mathrm{VNR}_{\mathcal{D},m}+\eta\,\Delta_{\mathrm{acc}}(k),\\
\Delta_{\mathrm{acc}}(k)&=10\log_{10}k\;\text{dB}.
\end{aligned}
\end{equation}
where $k$ is the number of received blocks in the session and $\eta=0.5$; fixed-length baselines use H-ARQ retransmissions without accumulation.

\begin{table}[t]
    \centering
    \caption{Representative urban V2X simulation parameters.}
    \label{tab:v2x_params}
    \resizebox{\columnwidth}{!}{%
    \begin{tabular}{@{}ll@{}}
        \toprule
        \textbf{Parameter} & \textbf{Value / model} \\
        \midrule
        Carrier / bandwidth & 5.9~GHz ITS band; 10~MHz \\
        Vehicle speed & 30--120~km/h (urban/highway) \\
        Fading model & Block Rayleigh, coherence time $T_c$ \\
        Erasure probability $\mathfrak{p}$ & 0.05--0.25 (sidelink occupancy) \\
        Eavesdropper SNR deficit $\Delta\gamma$ & 3--10~dB \\
        DF relay population $L$ & 4--12 vehicles \\
        Monte Carlo trials $N_{\mathrm{trials}}$ & $10^4$ per VNR point \\
        Rateless budget $M_{\max}$ & 8 packets \\
        Operating lattice dimension $n$ & 995 \\
        Fountain accumulation efficiency $\eta$ & 0.5 (Raptor only) \\
        Link-level decoder model & AWGN $\mathrm{FER}(\mathrm{VNR})$; Raptor boost in \eqref{eq:raptor_accum} \\
        \bottomrule
    \end{tabular}%
    }
    \vspace{-0.3cm}
\end{table}

Table~\ref{tab:v2x_params} summarizes the vehicular parameters used in Section~\ref{sec_simulation}.
The AWGN results in Figs.~\ref{fig:1001} and~\ref{fig:101} characterize the underlying lattice codec in the high-SNR, no-fading limit that is representative of short-range line-of-sight V2X links, whereas the fading--erasure evaluation quantifies delivery robustness under mobility-induced channel variability.
Fading and PLS curves employ the rateless-aware link-level model in Table~\ref{tab:v2x_params}; matched-length LDLC and LDPC lattices are compared under identical fading and erasure realizations in Fig.~\ref{fig:v2x_fading}.

\section{Rateless Raptor Lattice Construction}\label{sec_preliminary}
This section develops the physical-layer codec used in the V2X architecture of Section~\ref{sec_v2x}: fountain Raptor codes as Construction-D$'$ component codes and the associated multilevel lattice.
\subsection{Raptor Codes}
Raptor codes belong to the family of fountain codes, which are rateless and not constrained by fixed block sizes. A key feature of fountain codes, including Raptor codes, is that they generate an unlimited number of output symbols from the source message. These codes allow receiving devices to reconstruct the original data by collecting any set of encoding symbols greater than or equal to the number of source symbols.

 With rateless codes, each receiver can successfully decode the message with the required quality, regardless of the channel conditions. In fountain codes, the output symbols are generated randomly as a function of the entire message, and the transmitter has no knowledge of which output symbols have been received. These codes can approach the capacity of erasure channels without needing prior knowledge of the erasure probability. Specifically, a fountain code generates, for a given set of $k$ input symbols $(u_1, u_2, \dots, u_k)$, an unlimited stream of output symbols $\varphi_1, \varphi_2, \dots$. The input and output symbols can either be individual bits or, more generally, binary vectors of a specified length.

To recover the original message, the number of output symbols collected by the receiver must slightly exceed $k$, the number of input symbols. A decoding algorithm for a fountain code is one that can recover $k$ input symbols from any $m$ output symbols with high probability. For well-designed fountain codes over an erasure channel, $m$ is typically very close to $k$, and the decoding time is linear in $k$. The first class of such codes is known as LT codes.
In LT codes, each output symbol's degree \( d \) is determined based on the output degree distribution, which is a probability distribution over the set \(\{1, \dots, k\}\). Following this, \( d \) input symbols are randomly selected, and the value of the output symbol is obtained by appending the values of these selected input symbols. The encoder constructs a bipartite graph that connects the source symbols to the encoded symbols.
When the average degree \( \bar{d} \) is significantly smaller than \( k \), the graph becomes sparse, allowing for the consideration of an irregular low-density generator matrix. In such sparse graphs, the average degree of the output symbols approaches approximately \(\log_e{k}\), where $e$  is the base for natural logarithm. As a result, LT codes are classified as random linear fountain codes.

The decoding complexity of LT codes becomes generally infeasible as the code length increases. For an LT code, achieving an encoder and decoder with linear time complexity is difficult because the average output degree grows at least logarithmically with the number of input symbols. It has been shown that a capacity-achieving LT code exhibits an average degree that increases with the message size at a logarithmic rate, which makes the design of a linear-time encoder and decoder particularly challenging \cite{sara}.
To address this problem, Raptor codes were developed as a more practical version of LT codes. Raptor codes achieve linear-time encoding and decoding by concatenating an LT code with an outer code that compensates for the gaps in the LT code \cite{drshok}, \cite{dretesami}. As a result, Raptor codes represent an advanced class of random linear fountain codes.

A Raptor code is constructed by concatenating an LT code, with an average degree of approximately $\bar{d}$, with a high-rate pre-code. This design results in better performance compared to LT codes. Typically, an LDPC code is used as the pre-code. Some Raptor codes may be constructed using two-layer codes, such as an LDPC code and a Hamming code, as their pre-codes. The general idea is to employ a block code that has both an encoder and decoder with linear-time complexity. This pre-code recovers the original input symbols along with redundant symbols.

This paper adopts the method proposed in \cite{2019} to construct an LT code. This approach is based on heuristic parameters and aims to determine the optimal degrees for the parity bits, ${D_1, D_2, \dots, D_i, \dots}$, and their corresponding lengths, ${L_1, L_2, \dots, L_i, \dots}$. Consequently, in contrast to conventional Raptor and LT code construction techniques that rely on solving linear programming (LP) formulations, this method offers a more efficient approach by eliminating the computational burden of LP problem solving.
Drawing inspiration from \cite{2019}, which utilizes a degree set of $\mathcal{D} = \{12,6,4\}$ and corresponding $\mathcal{A} = \{0.27,0.146,1.39\}$ for
regular-graph-based LT code construction in AWGN channels, we began our design with $D_i= 4$ and $\alpha_i = L_i/N= 1.39$ for all $i$, guided by their heuristic. We also conducted simulations to fine-tune these initial parameters, ultimately selecting the values $D_i=D = 12$ and $\alpha_i =L_i/N= 0.27$ for all $i$ that provided the most effective trade-off between decoding efficiency and erasure resilience.

In this paper, QC-LDPC codes are employed as pre-codes. QC-LDPC codes are a subset of LDPC codes characterized by parity-check matrices composed of square  blocks that are either circulant permutation matrices or zero matrices. They offer several advantages, including low-complexity iterative decoding with good performance and straightforward encoding implementation due to the potential use of shift registers \cite{8my_thesis}. There are various methods to construct QC-LDPC codes, which can be categorized into two groups: algebraic methods and graph-based methods. After extensive experimentation with various families of QC-LDPC and LDPC codes, we observed optimal performance when using the QC-LDPC codes employed in the construction of QC-LDPC lattices \cite{malema2025, a}. The construction process involves specifying critical parameters, including girth \( g \), row weight \( w_r \), column weight \( w_c \), the size of each circulant \( \vartheta \), and a base matrix \( \mathbf{B} \), which is defined as a \( w_c \times w_r \) all-one matrix in our setting. Initially, identifying suitable parameter sets relied on a trial-and-error approach. However, we subsequently derived a set of rules to effectively determine appropriate pre-codes.
For our study,  we set \( g = 8 \) and optimize the parameters to maximize \( w_r \) while minimizing \( \vartheta \) for a fixed code length \( N = \vartheta \cdot w_r \). The column degree is typically set to $2$ or \( 3 \), although higher lengths may warrant a column degree of \( 4 \).
The QC-LDPC code construction algorithm systematically searches for shift values for each check and variable group, ensuring that the conditions for the desired girth are maintained.   It is important to note that for certain values of \( N \), it may not be possible to find a code with the desired parameter set. In such cases, we make slight adjustments to the value of \( N \) until an appropriate code is identified. For instance, in constructing a lattice of dimension \( 1000 \), we initially required a value of \( N \) around \( 700 \). However, finding a suitable code at precisely this value proved challenging. Instead, for \( N = 696 \) with \( w_r = 12 \), \( w_c = 3 \), and \( \vartheta = 58 \), we constructed a lattice of dimension \( 995 \). Similarly, for \( N = 708 \) with \( w_r = 12 \), \( w_c = 2 \), and \( \vartheta = 59 \), we achieved a lattice dimension of \( 1025 \). Notably, both lattices exhibit nearly identical error performance.
It is important to note that careful attention to these parameters is essential for ensuring the performance and reliability of the constructed Raptor codes and their corresponding lattices.
By employing  QC-LDPC code as the pre-code, we can leverage its structured nature to reduce the overall complexity of the Raptor code. This approach can result in faster encoding and decoding, as well as lower power consumption and memory usage.
Given the diverse range of applications for systematic codes, systematic Raptor codes are generally preferred over their non-systematic counterparts. A systematic LT code serves as a foundation for constructing systematic Raptor codes. This can be achieved by augmenting a non-systematic LT code generator matrix with an identity matrix, thereby producing a systematic LT code generator. In this study, systematic Raptor codes are utilized as subcodes for lattice construction.

\subsection{Lattices}\label{Parity_lattice}
A lattice is defined as a discrete additive subgroup of \(\mathbb{R}^n\). Analogous to linear codes, a lattice $\Lambda$ possesses a generator matrix, denoted as \(\mathbf{G}_{\Lambda}\). This relationship can be expressed as \(\mathbf{x} \in \Lambda\) if and only if \(\mathbf{x}= \mathbf{G}_{\Lambda}\mathbf{b}\), where \(\mathbf{b} \in \mathbb{Z}^n\). Indeed, lattice points are generated through integer linear combinations of the columns of the lattice generator matrix.
In the rateless setting adopted for V2X, each successfully collected fountain output induces a valid lattice point once the underlying Raptor component codewords are recovered; the effective code rate is therefore determined at the receiver rather than fixed at the encoder.

The \emph{volume} of a lattice \(\Lambda\) defined by a generator matrix \(\mathbf{G}_\Lambda\) is the volume of the fundamental region (or parallelepiped) formed by the basis vectors of the lattice. Mathematically, if we denote the matrix transpose by $(\cdot)^T$, the volume of a lattice \( \Lambda \) can be computed as
\(\mathrm{vol}(\Lambda) =\sqrt{\det(\mathbf{G}_\Lambda^T\mathbf{G}_\Lambda)}\), which equals \(|\det(\mathbf{G}_\Lambda)|\)
for full-rank lattices.

If \(\mathbf{U}\) is an $n \times n$ unimodular matrix, then the product \(\mathbf{G}_{\Lambda}\mathbf{U} \) defines an alternative generator matrix for the lattice \(\Lambda\). A unimodular matrix is characterized by having an integer determinant of \( \pm 1\), which implies that it preserves the integer structure of the vectors it operates upon and does not change the lattice volume.

    Construction-D$'$ is one of the methods used to construct a lattice by employing linear nested codes. In other words, a set of parity checks defining a family of codes is converted into lattice congruences through Construction-D$'$. In addition to the classic Construction-D$'$, there are two other definitions of this method. One of them is based on the congruences of subcodes' parity-check matrices \(\mathbf{H}_i\)'s, without requiring \(\mathbf{H}_i\)'s to be nested. In this generalized Construction-D$'$, \(\mathbf{H}_i\)'s  must satisfy a more general condition instead of being nested:
\begin{equation}\mathbf{H}_i \equiv \mathbf{F}_i \mathbf{H}_{i - 1} \pmod{2^i}, \quad 0 \leq i \leq a-1.\end{equation}
for some integer matrix \(\mathbf{F}_i\) to achieve an \(a\)-level lattice \cite{silva}.

The second definition of Construction-D$'$ is based on the lattice parity-check matrix, which is the inverse of the lattice generator matrix, that is, \(\mathbf{H}_{\Lambda} = \mathbf{G}_{\Lambda}^{-1}\) \cite{fan}.

The Construction-D$'$ version employed in this study integrates the two previously mentioned definitions. More specifically,  the matrices \(\mathbf{H}_i\)'s sequentially satisfy congruences, while the lattice points \(\mathbf{x}\) are obtained directly from \(\mathbf{H}_{\Lambda}\).
We consider a family of  binary nested  linear codes \( C_{0} \subseteq C_{1} \subseteq \cdots \subseteq C_{a} = \mathbb{F}_{2}^{n} \). In the classical formulation of Construction-D$'$, the parity-check matrices corresponding to this family of codes are  nested binary matrices: \( \mathbf{H}_{0} \sqsupseteq \mathbf{H}_{1} \cdots \sqsupseteq \mathbf{H}_{a} \), where the notation \(\mathbf{H}_{i-1} \sqsupseteq \mathbf{H}_{i}\) indicates that the rows of \(\mathbf{H}_{i}\) are a subset of the rows of \(\mathbf{H}_{i-1}\). However, in our approach, which is inspired by \cite{silva}, it is not a requirement for the parity-check matrices to be nested. The dimensions of the codes \(C_{i}\) are denoted as \(k_{i}\), with the corresponding rate defined as \(R_{i} = \frac{k_{i}}{n}\).

Let \( \widetilde{\mathbf{H}} \) be a unimodular matrix of size \( n \times n \) defined as follows:
\[
\widetilde{\mathbf{H}} = \begin{bmatrix}
\widetilde{\mathbf{h}}_{1} \\
\widetilde{\mathbf{h}}_{2} \\
\vdots \\
\widetilde{\mathbf{h}}_{n}
\end{bmatrix} = \begin{bmatrix}
\mathbf{h}_{1} \\
\vdots \\
\mathbf{h}_{k_{0}} \\
\mathbf{H}_{0}
\end{bmatrix},
\]
where this matrix includes the parity-check matrices of the nested subcodes, along with additional rows \( \mathbf{h}_{1} , \ldots, \mathbf{h}_{k_{0}} \) to ensure that it is a unimodular matrix. These additional rows do not contribute to the error correction process.

Let \( \mathbf{D} \) be a diagonal matrix with elements on the main diagonal defined as:
\[
d_{j,j} = 2^{-i}, \quad \text{for } k_{i-1} < j < k_{i},
\]
where \( k_{-1} = 0 \) and \( k_{a} = n \). Then, the check matrix of the Construction-D\('\) lattice is defined as:
\begin{equation}\mathbf{H}_{\Lambda} = \mathbf{D} \cdot \widetilde{\mathbf{H}}.\end{equation}
The points of this lattice consist of all vectors \( \mathbf{x} \in \mathbb{R}^n \) that yield an integer vector upon multiplication by \( \mathbf{H}_{\Lambda} \):
\[
\Lambda = \{ \mathbf{x}\in\mathbb{R}^n \mid \mathbf{H}_{\Lambda} \cdot \mathbf{x}=\mathbf{b} \in \mathbb{Z}^n \}.
\]

\section{Raptor-Lattice Encoding for V2X Broadcast}\label{sec_encoding}
As previously mentioned, a Raptor code is a compound code formed by employing a high-rate code as the pre-code and an LT code as the outer code. The rest of this section will discuss the construction of our employed Raptor codes in detail.
\subsection{The Raptor Code Design}
In this paper, we apply the approach outlined in \cite{2019}. The generator matrix of a Raptor code, denoted as \( \mathbf{G}_{\text{Raptor}} \), can be expressed as
$\mathbf{G}_{\text{Raptor}} = \mathbf{G}_{\text{out}} \mathbf{G}_{\text{in}},
$ where \( \mathbf{G}_{\text{out}} \) and \( \mathbf{G}_{\text{in}} \) are the generator matrices for the QC-LDPC part and the LT part, respectively. In other words, the generator matrix of the underlying Raptor code for creating a lattice consists of \( \mathbf{G}_{\text{in}} \) and \( \mathbf{G}_{\text{out}} \). The matrix \( \mathbf{G}_{\text{in}} \) is defined as follows:
\begin{equation*}
\mathbf{G}_{\text{in}}=\big[\mathbf{I}_{N}\;\;\mathbf{P}_{1}\;\;\mathbf{P}_{2}\;\cdots\;\mathbf{P}_{i}\;\cdots\big],
\end{equation*}
where \(\mathbf{I}_{N}\) is the identity matrix of size \( N \times N \). For each $i$, the matrix \( \mathbf{P}_{i} \) is a binary random matrix of size \( N \times L_{i} \), in which each column contains \( D_{i} \) ones. To produce the \( i^{th} \) group of parity bits of size \( L_{i} \times 1 \), denoted as \( \mathbf{r}_{i} \),  the matrix \( \mathbf{P}_{i} \) is utilized.

All parity-check nodes in the \(i^{th}\) group of parity nodes have a degree of \(D_i\) and are clustered rather than scattered. Each block \(\mathbf{P}_{i}\) is composed of several sub-blocks of size \(N \times \frac{N}{D_{i}}\). In the framework put forth by \cite{2019}, the construction process is designed such that precisely one \(``1"\) is placed in each row of every sub-block, thereby preventing the formation of cycles among the bits within the Tanner graph of each sub-block.
The objective in their design is to randomly place \(1\)s within each sub-block while satisfying two constraints: 1) Having exactly one \(1\) per row, and 2) Having \(D_i\) \(1\)'s per column.
If a number corresponds to a valid location (satisfying both constraints), a \(1\) is placed there. Otherwise, it is disregarded, and the next number is checked.

To this end, first, we set \(L_{i} = \alpha_{i} N\), and then the primary objective is to find optimized values for \(D_{i}\)'s and \(\alpha_{i}\)'s. The degree set $\mathcal{D}=\{D_{1}, D_{2}, \ldots, D_{i}, \ldots\}$ and the set \(\mathcal{A}=\{\alpha_{1}, \alpha_{2}, \ldots, \alpha_{i}, \ldots\}\) are significant for this method of constructing LT codes, and based on them, the generator matrix \(\mathbf{G}_{\textrm{in}}\) is designed. The strategy for determining these two sequences is explained in \cite{2019}. Finally, each bit in the \(i^{th}\) group of parity bits \(\mathbf{r}_{i}\) has a degree \(D_{i}\) from the set \(\{D_{1}, D_{2}, \ldots, D_{i}, \ldots\}\), which is a non-increasing sequence of positive integers.
Algorithm~\ref{alg:alg1} outlines the procedure for deriving the generator matrix of LT codes from the given sequences.

\begin{algorithm}[tb]
\caption{Generator Matrix of LT Codes}
\label{alg:alg1}
\begin{algorithmic}[1] 
\State \textbf{Input:} \(N, \mathcal{A} = \{ {\alpha _1}, {\alpha _2}, \ldots \}, \mathcal{D} = \{ {D_1}, {D_2}, \ldots \}\)
\State \(N\) is equal to \(k\), the number of rows in \(\mathbf{G}_{\mathrm{LT}}\), and the number of columns in \(\mathbf{G}_{\mathrm{QC-LDPC}}\)
\State \textbf{Output:} \({\mathbf{G}_{LT}} \in \{ 0,1\}^{k \times n}\)
\For{$i = 1$ \textbf{to} $\infty$}
   \State \(\mathbf{P}_i \in \{ 0,1\}^{N \times (\alpha_i \cdot N)}\)
   \For{$j = 1$ \textbf{to} $\left\lfloor \frac{\alpha_i N}{\frac{N}{D_i}} \right\rfloor$}
       \State Block\(-k_j = \{ 0,1\}^{N \times \left\lfloor \frac{N}{D_i} \right\rfloor}\)
       \For{$q = 1$ \textbf{to} $\frac{N}{D_i}$}
           \While{$|\{\mathrm{Block-}{k_j}(:,q)\} \neq 0| < D_i$}
               \State \textbf{select randomly} \(l: 0 < l \leq N\)
               \If{$|\{\mathrm{Block-}{k_j}(l,:)\} \neq 0| = 0$}
                   \State \(\mathrm{Block-}{k_j}(l,q) = 1\)
               \EndIf
           \EndWhile
       \EndFor
    \EndFor
\EndFor
\end{algorithmic}
\end{algorithm}

\subsection{Encoding Construction-D$'$ Lattices}
This section adheres to the methodology outlined in \cite{fan} for encoding Construction-D$'$ Lattices. Consider an integer message sequence $\mathbf{b}$, consisting of information vectors $\mathbf{u}_{i} \in \mathbb{F}_{2}^{k_{i}}$ of code $C_{i}$, where $0 \leq i \leq a-1$ is provided, alongside an integer vector $\mathbf{z} \in \mathbb{Z}^{n}$. This encoding constitutes a mapping wherein the message sequence $\mathbf{b}$ is represented as a lattice point $\mathbf{x}$. As previously stated, for an integer vector $\mathbf{b} \in \mathbb{Z}^{n}$, its corresponding lattice point  $\mathbf{x}$ will be obtained by $\mathbf{x} = \mathbf{G}_{\Lambda} \cdot \mathbf{b}$.

If we consider
${\mathbf{u}_{i}}^{\prime}$
as the zero-padded form of
$\mathbf{u}_{i}$ by adding $n-k_i$ zeros to accommodate a vector of size
$n \times 1$, then the message sequence $\mathbf{b}$ can be written as follows:
\begin{equation}\label{eq1}
\mathbf{b}=\mathbf{D}\!\cdot\!\Big(\sum\limits_{i=0}^{a-1}2^{i}\mathbf{u}_{i}^{\prime}+2^{a}\mathbf{z}\Big).
\end{equation}
On the other hand, for a point
$\mathbf{x}$
in a Construction-D$'$ lattice, we can decompose $\mathbf{x}$ as $\mathbf{x} = \sum\limits_{i=0}^{a} 2^{i} \mathbf{x}_{i}$,
where the components $\mathbf{x}_{i}$'s
are not necessarily binary and each
$\mathbf{x}_{i}$ is congruent modulo 2 to a codeword ${{\mathbf{\widetilde x}}_i} \in {C_i}$.
Given that the components $\mathbf{x}_{i}$'s are dependent on  $\mathbf{u}_{i}$'s, it is possible to demonstrate the correspondence between a lattice point $\mathbf{x}$ and the information vectors $\mathbf{u}_{i}$'s as will be detailed in the subsequent exposition.
Indeed, since
$\mathbf{G}_{\Lambda} \cdot \mathbf{b} = \mathbf{x}$,
and based on the definition from \cite{fan},
$\mathbf{G}_{\Lambda} = \mathbf{H}_{\Lambda}^{-1}$, we can write
$
\mathbf{H}_{\Lambda} \cdot \mathbf{x} = \mathbf{b}.
$
Since
$\mathbf{H}_{\Lambda} = \mathbf{D} \cdot \widetilde{\mathbf{H}}$
it follows that
$
\widetilde{\mathbf{H}} \cdot \mathbf{x} = \mathbf{D}^{-1} \cdot \mathbf{b}.
$
Now, based on (\ref{eq1}) and decomposition of $\mathbf{x}$, we have:
\begin{equation}
\begin{aligned}
&\widetilde{\mathbf{H}} \cdot \left( \sum\limits_{i=0}^{a} 2^{i} \mathbf{x}_{i} \right)\\
&\qquad= \sum\limits_{i=0}^{a-1} 2^{i} \mathbf{u}_{i}^{\prime} + 2^{a} \mathbf{z}.
\end{aligned}
\end{equation}
Therefore, the lattice components
$\mathbf{x}_{i}$'s  in $\mathbb{Z}^{n}$
are retrieved by solving the following equations:
\begin{align*}
\widetilde{\mathbf{H}} \cdot \mathbf{x}_{i} &= \mathbf{u}_{i}^{\prime}, \;\;\;\;\; 0 \leq i \leq a-1, \\
\widetilde{\mathbf{H}} \cdot \mathbf{x}_{a} &= \mathbf{z}.
\end{align*}

\section{Multistage Decoding at V2X Receivers}\label{Sec_decoding}
In this section, we focus on the decoding process of the Raptor codes, which involves both the QC-LDPC and LT components. The decoding can be effectively represented and handled using a modified Tanner graph structure
\cite{IET}. By adapting the Tanner graph, we are able to integrate both the QC-LDPC and LT codes into a unified graph structure that enables efficient iterative decoding. This modification enables the utilization of parity-check equations as termination conditions for the decoding algorithm, thereby enhancing the overall performance.
\subsection{Modified Tanner Graph}
Through modification of the representing graph, we can present the Tanner graph for both component codes (QC-LDPC and LT) with the same structure. Parity-check equations in iterative decoding algorithms serve as the stopping criterion. Therefore, by representing the LT component through check and bit nodes, it becomes feasible to employ parity-check equations for the entire decoding process of the Raptor component code. Consequently, it is sufficient to represent the LT component with check nodes and bit nodes after it is combined with the Tanner graph of the QC-LDPC component. This results in the unified Raptor Tanner graph. In this manner, the desired iterative decoding algorithm will be able to operate on both components of the Raptor code.

\begin{figure}[!t]
    \centering
    \resizebox{\columnwidth}{!}{
    \definecolor{ubqqys}{rgb}{0.29411764705882354,0,0.5098039215686274}
\definecolor{qqffqq}{rgb}{0,1,0}
\definecolor{ffqqqq}{rgb}{1,0,0}
\definecolor{qqqqcc}{rgb}{0,0,0.8}
\begin{tikzpicture}[line cap=round,line join=round,>=triangle 45,x=0.55cm,y=0.55cm]
\clip(-14.198095238095242,-5.093333333333349) rectangle (15.925714285714276,5);
\draw [line width=1.2pt,color=ubqqys] (-12,4)-- (-12,0);
\draw [line width=1.2pt,color=ubqqys] (-6,4)-- (-6,0);
\draw [line width=1.2pt] (-12,0)-- (-12,-4);
\draw [line width=1.2pt] (-6,0)-- (-6,-4);
\draw [line width=1.2pt] (-2,0)-- (-2,-4);
\draw [line width=1.2pt] (0,0)-- (0,-4);
\draw [line width=1.2pt,color=ubqqys] (-6,4)-- (0,0);
\draw [line width=1.2pt,color=ubqqys] (-6,4)-- (-2,0);
\draw (6.5,-3.4) node[anchor=north west] {\huge$\cdots$};
\draw (-11,4.65714285714291) node[anchor=north west] {\huge$\cdots$};
\draw (-11,0.5209523809523806) node[anchor=north west] {\huge$\cdots$};
\draw (-11,-3.4) node[anchor=north west] {\huge$\cdots$};
\draw (-5,-3.4) node[anchor=north west] {\huge$\cdots$};
\draw (-5,0.5419047619047617) node[anchor=north west] {\huge$\cdots$};
\draw (0.9152380952380885,-3.4) node[anchor=north west] {\huge$\cdots$};
\draw [line width=1.2pt,color=ubqqys] (-8,4)-- (-8,0);
\draw [line width=1.2pt] (-8,0)-- (-8,-4);
\draw [line width=1.2pt,color=ubqqys] (-8,4)-- (0,0);
\draw [line width=1.2pt,color=ubqqys] (-9.992380952380953,3.8266666666666707)-- (-4.5857142857142904,0.3);
\draw [line width=1.2pt] (-3.6857142857142904,0)-- (6,-4);
\draw [line width=1.2pt] (-4.9009523809523845,0)-- (4,-4);
\draw [line width=1.2pt] (0,0)-- (6,-4);
\draw [line width=1.2pt] (0,0)-- (4,-4);
\draw [line width=1.2pt] (-10.013333333333334,-0.1542857142857156)-- (4,-4);
\draw (-12.47619047619048,5.21428571428572) node[anchor=north west] {$u_1$};
\draw (-8.653333333333337,5.130476190476196) node[anchor=north west] {$u_{K-1}$};
\draw (-6.46285714285714,5.172380952380958) node[anchor=north west] {$u_K$};
\draw (-2.596190476190482,4.397142857142862) node[anchor=north west] {Information Bits};
\draw (2.3485714285714216,0.24857142857142792) node[anchor=north west] {LDPC Intermediate Bits};
\draw (7.293333333333325,-4) node[anchor=north west] {Raptor Bits};
\draw (-13.567619047619051,1) node[anchor=north west] {$m_1$};
\draw (-10.263809523809529,1) node[anchor=north west] {$m_{K-1}$};
\draw (-7.44476190476191,1) node[anchor=north west] {$m_{K}$};
\draw (-4.5723809523809584,1) node[anchor=north west] {$m_{N-1}$};
\draw (0.0857142857142792,1) node[anchor=north west] {$m_N$};
\draw (-12.47619047619048,-4.235238095238102) node[anchor=north west] {$\varphi_1$};
\draw (-8.653333333333337,-4.1723809523809585) node[anchor=north west] {$\varphi_{K-1}$};
\draw (-6.462857142857148,-4.235238095238102) node[anchor=north west] {$\varphi_K$};
\draw (-2.681904761904768,-4.1514285714285775) node[anchor=north west] {$\varphi_{N-1}$};
\draw (-0.42857142857143503,-4.1514285714285775) node[anchor=north west] {$\varphi_N$};
\draw (3.331428571428564,-4.1723809523809585) node[anchor=north west] {$\varphi_{n-1}$};
\draw (5.563809523809516,-4.1723809523809585) node[anchor=north west] {$\varphi_n$};
\begin{scriptsize}
\draw [fill=qqqqcc] (-12,4) circle (4.5pt);
\draw [fill=qqqqcc] (-6,4) circle (4.5pt);
\draw [fill=ffqqqq] (-12,0) circle (4.5pt);
\draw [fill=ffqqqq] (-6,0) circle (4.5pt);
\draw [fill=ffqqqq] (-2,0) circle (4.5pt);
\draw [fill=ffqqqq] (0,0) circle (4.5pt);
\draw [fill=qqffqq] (-12,-4) circle (4.5pt);
\draw [fill=qqffqq] (-6,-4) circle (4.5pt);
\draw [fill=qqffqq] (-2,-4) circle (4.5pt);
\draw [fill=qqffqq] (0,-4) circle (4.5pt);
\draw [fill=qqffqq] (4,-4) circle (4.5pt);
\draw [fill=qqffqq] (6,-4) circle (4.5pt);
\draw [fill=qqqqcc] (-8,4) circle (4.5pt);
\draw [fill=ffqqqq] (-8,0) circle (4.5pt);
\draw [fill=qqffqq] (-8,-4) circle (4.5pt);
\end{scriptsize}
\end{tikzpicture}
    }
\caption{Factor graph for a systematic Raptor code. The first $K$ bit nodes are mapped to $N=k$ intermediate nodes, which are then considered as input nodes for the LT code. The last layer corresponds to the encoding nodes of the Raptor code. This Tanner graph is associated with the code $C_i$, one of the subcodes in the Construction-D$'$ lattice. The message sequence $\mathbf{b}$ consists of
    $\mathbf{u}_i \in \mathbb{F}_2^{k_i}$
from the code $C_i$, for $0 \leq i \leq a-1$,
and an integer vector
$\mathbf{z} \in \mathbb{Z}^n$.
In other words, the elements of $\mathbf{b}$ are derived from the vectors $\mathbf{u}_i$ of the codes $C_i$ and the vector $\textbf{z}$.}
    \label{f}
    \vspace{-0.4cm}
\end{figure}
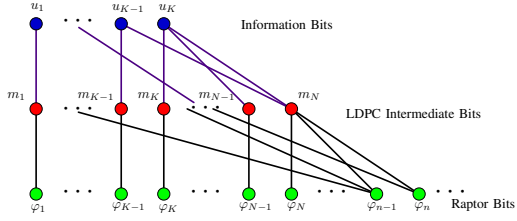

According to the systematic specifications of our Raptor codes, as shown in Fig. \ref{f}, the systematic part of the coded bits
$\varphi_{i}$'s, for $i \leq k \,(= N)$, is the same as the intermediate bits $m_{i}$'s. This means
$\varphi_{i} = m_{i}$, for $i=1,\ldots,k$.
Thus, the parity parts of the coded bits are given by:
\[
\varphi_{j} = \underset{i \in \chi_{j}}{\oplus} m_{i} = \underset{i \in \chi_{j}}{\oplus} \varphi_{i}, \;\;\; j > k,
\]
where $\oplus$ denotes the binary addition and ${{\chi }_{j}}$
 is the set of neighboring nodes of the coded node
 $\varphi_j$.
To modify the Tanner graph, the encoded bits of the LT component must first be converted into bit nodes, followed by the addition of check nodes denoted as \( CH_{t} \). The parity-check equations can be expressed as:
\begin{equation*}
\begin{aligned}
CH_{t}&:=\underset{i\in\mathcal{T}_{t}}{\oplus}\varphi_{i}
=\varphi_{j}\underset{i\in\chi_{j}}{\oplus}\varphi_{i},\\
&\qquad 1\leq t\leq n-k,\; i\in\chi_{j}.
\end{aligned}
\end{equation*}
where \(\mathcal{T}_{t} = \{ t \mid t = j - k, \; j > k \}\).
By applying a soft iterative decoding algorithm to the modified Tanner graph of the LT codes, more reliable soft information can be obtained, leading to enhanced decoding performance \cite{18IET}. This improved performance of the LT component contributes directly to the overall enhancement of the Raptor code's effectiveness, as demonstrated in \cite{IET}.

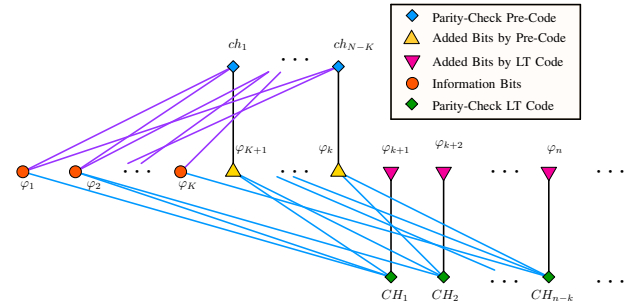
\begin{figure}[t]
    \centering
    \definecolor{ffcctt}{rgb}{1,0.8,0.2}
\definecolor{ffeacd}{rgb}{1,0.9176470588235294,0.803921568627451}
\definecolor{zzttff}{rgb}{0.6,0.2,1}
\definecolor{fuqqzz}{rgb}{0.9568627450980393,0,0.6}
\definecolor{ffdxqq}{rgb}{1,0.8431372549019608,0}
\definecolor{ffvvqq}{rgb}{1,0.3333333333333333,0}
\definecolor{qqzzff}{rgb}{0,0.6,1}
\definecolor{qqzzqq}{rgb}{0,0.6,0}
\resizebox{\columnwidth}{!}{
\begin{tikzpicture}[line cap=round,line join=round,>=triangle 45,x=0.70cm,y=0.70cm]
\clip(-12.8092952380952377,-8.87619047619049) rectangle (12.297142857142838,2.408095238095237);
\fill[line width=1.2pt,color=ffeacd,fill=ffeacd,fill opacity=0.1] (2,-2) -- (1.99867,2.36) -- (9.2,2.35638) -- (9.19695,-2) -- cycle;
\draw (-8.418857142857146,-3.6) node[anchor=north west] {\huge$\cdots$};
\draw (-2.426476190476197,-3.6) node[anchor=north west] {\huge$\cdots$};
\draw (-2.403428571428578,0.6278095238095184) node[anchor=north west] {\huge$\cdots$};
\draw (5.571047619047607,-3.6) node[anchor=north west] {\huge$\cdots$};
\draw (1.4542857142857047,-8.3) node[anchor=north west] {$CH_1$};
\draw (-4.362476190476196,1.227047619047614) node[anchor=north west] {$ch_{1}$};
\draw (-0.3521904761904844,1.250095238095233) node[anchor=north west] {$ch_{N-K}$};
\draw (-12.267809523809523,-4.19695238095239) node[anchor=north west] {$\varphi_1$};
\draw (-9.73961904761905,-4.19695238095239) node[anchor=north west] {$\varphi_{2}$};
\draw (-6.252380952380957,-4.19695238095239) node[anchor=north west] {$\varphi_K$};
\draw (-4.092190476190481,-2.8) node[anchor=north west] {$\varphi_{K+1}$};
\draw (-0.9283809523809602,-2.8) node[anchor=north west] {$\varphi_k$};
\draw (1.468571428571419,-2.8) node[anchor=north west] {$\varphi_{k+1}$};
\draw (7.760571428571415,-2.8) node[anchor=north west] {$\varphi_n$};
\draw (9.581333333333319,-3.6) node[anchor=north west] {\huge$\cdots$};
\draw (3.496761904761894,-2.6449523809523887) node[anchor=north west] {$\varphi_{k+2}$};
\draw [line width=1.2pt] (-4,0)-- (-4,-4);
\draw [line width=1.2pt] (0,0)-- (0,-4);
\draw (3.373714285714275,-8.3) node[anchor=north west] {$CH_2$};
\draw (7.153142857142844,-8.3) node[anchor=north west] {$CH_{n-k}$};
\draw [line width=1.2pt] (2,-4)-- (2,-8);
\draw [line width=1.2pt] (4,-4)-- (4,-8);
\draw [line width=1.2pt] (8,-4)-- (8,-8);
\draw [line width=1.2pt,color=zzttff] (-4,0)-- (-12,-4);
\draw [line width=1.2pt,color=zzttff] (-4,0)-- (-10,-4);
\draw [line width=1.2pt,color=zzttff] (0,0)-- (-8.004,-3.6820952380952465);
\draw [line width=1.2pt,color=zzttff] (-6,-4)-- (-2.2038095238095234,-0.21714285714286322);
\draw [line width=1.2pt,color=zzttff] (-7.4969523809523775,-3.6820952380952465)-- (-2.6569523809523803,-0.2019047619047679);
\draw [line width=1.2pt,color=zzttff] (-12,-4)-- (0,0);
\draw [line width=1.2pt,color=zzttff] (-10,-4)-- (-2.6569523809523803,-0.2019047619047679);
\draw [line width=1.2pt,color=qqzzff] (-4,-4)-- (2,-8);
\draw [line width=1.2pt,color=qqzzff] (-4,-4)-- (4,-8);
\draw [line width=1.2pt,color=qqzzff] (0,-4)-- (4,-8);
\draw [line width=1.2pt,color=qqzzff] (0,-4)-- (8,-8);
\draw (5.548,-7.8) node[anchor=north west] {\huge$\cdots$};
\draw (9.581333333333319,-7.8) node[anchor=north west] {\huge$\cdots$};
\draw [line width=1.2pt,color=qqzzff] (-2.334285714285718,-4.1891428571428655)-- (5.939809523809513,-7.738476190476201);
\draw [line width=1.2pt,color=qqzzff] (-1.712,-4.1891428571428655)-- (8,-8);
\draw [line width=1.2pt,color=qqzzff] (-12,-4)-- (2,-8);
\draw [line width=1.2pt,color=qqzzff] (-10,-4)-- (2,-8);
\draw [line width=1.2pt,color=qqzzff] (-10,-4)-- (4,-8);
\draw [line width=1.2pt,color=qqzzff] (-6,-4)-- (8,-8);
\draw [line width=1.2pt,color=ffeacd] (2,-2)-- (1.99867,2.36);
\draw [line width=1.2pt,color=ffeacd] (1.99867,2.36)-- (9.2,2.35638);
\draw [line width=1.2pt,color=ffeacd] (9.2,2.35638)-- (9.19695,-2);
\draw [line width=1.2pt,color=ffeacd] (9.19695,-2)-- (2,-2);
\draw (3.4,2.2337142857142814) node[anchor=north west] {Parity-Check Pre-Code};
\draw (3.4,1.450095238095233) node[anchor=north west] {Added Bits by Pre-Code};
\draw (3.4,0.5434285714285659) node[anchor=north west] {Added Bits by LT Code};
\draw (3.4,-0.3093333333333395) node[anchor=north west] {Information Bits};
\draw (3.4,-1.1238095238095305) node[anchor=north west] {Parity-Check LT Code};
\draw [line width=1.2pt] (1.99867,2.36)-- (9.2,2.35638);
\draw [line width=1.2pt] (9.2,2.35638)-- (9.19695,-2);
\draw [line width=1.2pt] (9.19695,-2)-- (2,-2);
\draw [line width=1.2pt] (2,-2)-- (1.99867,2.36);
\begin{scriptsize}
\draw [fill=qqzzqq] (2,-8) ++(-4.5pt,0 pt) -- ++(4.5pt,4.5pt)--++(4.5pt,-4.5pt)--++(-4.5pt,-4.5pt)--++(-4.5pt,4.5pt);
\draw [fill=qqzzqq] (8,-8) ++(-4.5pt,0 pt) -- ++(4.5pt,4.5pt)--++(4.5pt,-4.5pt)--++(-4.5pt,-4.5pt)--++(-4.5pt,4.5pt);
\draw [fill=qqzzff] (-4,0) ++(-4.5pt,0 pt) -- ++(4.5pt,4.5pt)--++(4.5pt,-4.5pt)--++(-4.5pt,-4.5pt)--++(-4.5pt,4.5pt);
\draw [fill=qqzzff] (0,0) ++(-4.5pt,0 pt) -- ++(4.5pt,4.5pt)--++(4.5pt,-4.5pt)--++(-4.5pt,-4.5pt)--++(-4.5pt,4.5pt);
\draw [fill=ffvvqq] (-12,-4) circle (4.5pt);
\draw [fill=ffvvqq] (-6,-4) circle (4.5pt);
\draw [fill=ffdxqq,shift={(-4,-4)}] (0,0) ++(0 pt,6.75pt) -- ++(5.84567147554496pt,-10.125pt)--++(-11.69134295108992pt,0 pt) -- ++(5.84567147554496pt,10.125pt);
\draw [fill=ffdxqq,shift={(0,-4)}] (0,0) ++(0 pt,6.75pt) -- ++(5.84567147554496pt,-10.125pt)--++(-11.69134295108992pt,0 pt) -- ++(5.84567147554496pt,10.125pt);
\draw [fill=fuqqzz,shift={(2,-4)},rotate=180] (0,0) ++(0 pt,6.75pt) -- ++(5.84567147554496pt,-10.125pt)--++(-11.69134295108992pt,0 pt) -- ++(5.84567147554496pt,10.125pt);
\draw [fill=fuqqzz,shift={(8,-4)},rotate=180] (0,0) ++(0 pt,6.75pt) -- ++(5.84567147554496pt,-10.125pt)--++(-11.69134295108992pt,0 pt) -- ++(5.84567147554496pt,10.125pt);
\draw [fill=qqzzqq] (4,-8) ++(-4.5pt,0 pt) -- ++(4.5pt,4.5pt)--++(4.5pt,-4.5pt)--++(-4.5pt,-4.5pt)--++(-4.5pt,4.5pt);
\draw [fill=ffvvqq] (-10,-4) circle (4.5pt);
\draw [fill=fuqqzz,shift={(4,-4)},rotate=180] (0,0) ++(0 pt,6.75pt) -- ++(5.84567147554496pt,-10.125pt)--++(-11.69134295108992pt,0 pt) -- ++(5.84567147554496pt,10.125pt);
\draw [fill=qqzzff] (2.8053333333333232,1.941523809523805) ++(-4.5pt,0 pt) -- ++(4.5pt,4.5pt)--++(4.5pt,-4.5pt)--++(-4.5pt,-4.5pt)--++(-4.5pt,4.5pt);
\draw [fill=ffcctt,shift={(2.782285714285704,1.0426666666666615)}] (0,0) ++(0 pt,6.75pt) -- ++(5.84567147554496pt,-10.125pt)--++(-11.69134295108992pt,0 pt) -- ++(5.84567147554496pt,10.125pt);
\draw [fill=fuqqzz,shift={(2.8053333333333232,0.236)},rotate=180] (0,0) ++(0 pt,6.75pt) -- ++(5.84567147554496pt,-10.125pt)--++(-11.69134295108992pt,0 pt) -- ++(5.84567147554496pt,10.125pt);
\draw [fill=ffvvqq] (2.8127619047618944,-0.6398095238095302) circle (4.5pt);
\draw [fill=qqzzqq] (2.789714285714275,-1.4234285714285781) ++(-4.5pt,0 pt) -- ++(4.5pt,4.5pt)--++(4.5pt,-4.5pt)--++(-4.5pt,-4.5pt)--++(-4.5pt,4.5pt);
\end{scriptsize}
\end{tikzpicture}
}
    \caption{The Raptor Tanner graph is constructed by combining the Tanner graph of the LT code with the Tanner graph of the QC-LDPC component, resulting in a mixed graph that corresponds to the parity-check matrix of the Raptor code. In this graph, the nodes \( ch_1, \ldots, ch_{N-K} \) represent check nodes associated with the QC-LDPC part, while the nodes \( CH_1, \ldots, CH_{n-k} \) correspond to check nodes from the LT part.
    }
    \label{jj}
    \vspace{-0.3cm}
\end{figure}

As previously stated, our Raptor code is constructed in a systematic manner; consequently, the parity-check matrix of our Raptor codes has the following form:

\begin{equation}\label{Raptor_parity_check}
\resizebox{0.95\columnwidth}{!}{$
\mathbf{H}_{\text{Raptor}} = \begin{bNiceMatrix}[margin]
\Block[borders={bottom,right}]{1-1}{{{\mathbf{P}}^T_{\text{LDPC}}}} & \Block[borders={bottom,right}]{1-1}{\mathbf{I}_{(N - K)}} & \Block[borders={bottom}]{1-1}{\mathbf{0}_{(N-K)\times(n-N)}} \\
\Block[borders={right}]{1-2}{{{\mathbf{P}}^T_{\text{LT}}}} & &\Block[borders={}]{1-1}{\mathbf{I}_{(n - k)}} \\
\end{bNiceMatrix}_{(n - K) \times n}
$}
\end{equation}
where the upper left block matrix represents the parity-check matrix of the underlying LDPC code $\mathbf{H}_0$. Under the systematic encoding assumption, we have $\varphi_j = m_j = u_j$ for $1 \leq j \leq K$, and $\varphi_i = m_i$ for $K + 1 \leq i \leq N(=k)$.
Based on this feature, we can combine the Tanner graphs of the QC-LDPC and LT codes. Evidently, the $\mathbf{H}_{\text{Raptor}}$ matrix corresponds to the Tanner graph presented in Fig.~\ref{jj}. The presence of identity matrices in \eqref{Raptor_parity_check} leads to the formation of black edges within the Tanner graph of Fig.~\ref{jj}, whereas the zero matrix guarantees the absence of connections between the QC-LDPC component check nodes and the nodes $\varphi_{k + 1}, \cdots, \varphi_{n}$ in $\mathbf{H}_{\text{Raptor}}$. In this graph, the initial layer corresponds to the  check nodes of the QC-LDPC Tanner graph, the subsequent layer is associated with the output of the Raptor code, and the final layer comprises the check nodes related to the Tanner graph of the LT code.

We also emphasize that the parity-check matrix described in \eqref{Raptor_parity_check} corresponds to one of the underlying codes, not directly to the lattice itself. To construct the parity-check matrix of our Construction-D$'$ lattice, at least two such Raptor matrices, such as $\mathbf{H}_0$ and $\mathbf{H}_1$, are required. We then follow the method proposed in Section~\ref{Parity_lattice} to obtain the parity-check matrix of the lattice.
It is important to mention that incorporating girth-6 QC-LDPC and LT codes into the parity-check matrix of a Raptor code inherently yields a final Tanner graph with a girth of 4, leading to a poor error performance. To address this issue, we execute an additional modification step on our Tanner graph to eliminate the short cycles. Specifically, we implement Algorithm~\ref{girth4} on matrices $\mathbf{H}_0$ and $\mathbf{H}_1$.
\begin{algorithm}[t]
    \caption{Removing Girth-4 Cycles from $\mathbf{H}$}
    \label{girth4}
    \textbf{Input:} Parity-Check Matrix \( \mathbf{H}\) \\
    \textbf{Output:} Modified Parity-Check Matrix $\mathbf{H}'$
    \begin{algorithmic}[1]
    \State    $\mathbf{H}' \gets \mathbf{H}$
        \For{$r_1= 1$ to \text{rows}$(\mathbf{H})- 1$ }
            \For{$c_1 = 1$ to \text{cols}$(\mathbf{H})- 1$ }
                \If{$\mathbf{H}(r_1, c_1) == 1$}
                    \For{$c_2 = c_1 + 1$ to \text{cols}$(\mathbf{H})$}
                        \If{$\mathbf{H}(r_1, c_2) == 1$}
                            \For{$r_2 = r_1 + 1$ to \text{rows}$(\mathbf{H})$}
                                \If{$\mathbf{H}(r_2, c_2) == 1$ }
                                    \If{$\mathbf{H}(r_2, c_1) == 1$ }
                                        \State $\mathbf{H}'(r_2, c_1)$ \(\gets 0\)
                                    \EndIf
                                \EndIf
                            \EndFor
                        \EndIf
                    \EndFor
                \EndIf
            \EndFor
        \EndFor
    \end{algorithmic}
\end{algorithm}

\begin{example}
To demonstrate the effect of Algorithm~\ref{girth4}, \figurename~\ref{Tanner1} and \figurename~\ref{Tanner2} compare the Tanner graph of a random code before and after the girth-4 cycle removal process.
  \begin{figure}[t]
    \centering
  \resizebox{0.95\columnwidth}{!}{\begin{tikzpicture}
\tikzset{variable_node_style/.style={circle,draw=blue,fill=blue!30!white, inner sep = 0.1pt}};
\tikzset{check_node_style/.style={regular polygon,regular polygon sides=4,draw=green!50!black,fill=green!30!white, inner sep = 0.1pt}};
\tikzset{edge_style/.style={draw=black,line width=1pt }};
\tikzset{edge_style2/.style={draw=red,line width=1pt }};
\node [variable_node_style, label=below:{$\varphi_{1}$}] (x1) at (-2.416590,2) {$=$};
\node [variable_node_style, label=below:{$\varphi_{2}$}] (x2) at (-1.467350,2) {$=$};
\node [variable_node_style, label=below:{$\varphi_{3}$}] (x3) at (-0.463346,2) {$=$};
\node [variable_node_style, label=below:{$\varphi_{4}$}] (x4) at (0.476767,2) {$=$};
\node [variable_node_style, label=below:{$\varphi_{5}$}] (x5) at (1.426007,2) {$=$};
\node [variable_node_style, label=below:{$\varphi_{6}$}] (x6) at (2.256592,2) {$=$};
\node [variable_node_style, label=below:{$\varphi_{7}$}] (x7) at (2.968522,2) {$=$};
\node [variable_node_style, label=below:{$\varphi_{8}$}] (x8) at (3.762598,2) {$=$};
\node [variable_node_style, label=below:{$\varphi_{9}$}] (x9) at (4.529291,2) {$=$};
\node [variable_node_style, label=below:{$\varphi_{10}$}] (x10) at (5.204712,2) {$=$};

\node [check_node_style,label=above:{$ch_{1}$}] (c1) at (-1.467350,4.17) {$+$};
\node [check_node_style,label=above:{$ch_{2}$}] (c2) at (-0.527237,4.17) {$+$};
\node [check_node_style,label=above:{$ch_{3}$}] (c3) at (0.531530,4.17) {$+$};
\node [check_node_style,label=above:{$ch_{4}$}] (c4) at (1.581171,4.17) {$+$};
\node [check_node_style,label=above:{$ch_{5}$}] (c5) at (2.603429,4.17) {$+$};
\node [check_node_style,label=above:{$ch_{6}$}] (c6) at (3.735216,4.17) {$+$};

\path [edge_style] (x1) -- (c1);
\path [edge_style] (x1) -- (c3);
\path [edge_style] (x1) -- (c5);
\path [edge_style] (x2) -- (c2);
\path [edge_style2] (x2) -- (c3);
\path [edge_style2] (x2) -- (c5);
\path [edge_style] (x2) -- (c6);
\path [edge_style] (x3) -- (c5);
\path [edge_style] (x3) -- (c6);
\path [edge_style] (x4) -- (c1);
\path [edge_style] (x4) -- (c2);
\path [edge_style] (x4) -- (c3);
\path [edge_style] (x4) -- (c5);
\path [edge_style] (x4) -- (c6);
\path [edge_style] (x5) -- (c2);
\path [edge_style2] (x5) -- (c3);
\path [edge_style] (x5) -- (c4);
\path [edge_style2] (x5) -- (c5);
\path [edge_style] (x5) -- (c6);
\path [edge_style] (x6) -- (c5);
\path [edge_style] (x7) -- (c6);
\path [edge_style] (x8) -- (c3);
\path [edge_style] (x8) -- (c4);
\path [edge_style] (x9) -- (c3);
\path [edge_style] (x9) -- (c5);
\path [edge_style] (x9) -- (c6);
\path [edge_style] (x10) -- (c6);
\end{tikzpicture}
  }
    \caption{Initial Tanner graph structure showing short cycles before the application of Algorithm~\ref{girth4} for girth-4 cycle removal. Red edges indicate a representative cycle of length four.}
    \label{Tanner1}
    \vspace{-0.3cm}
\end{figure}
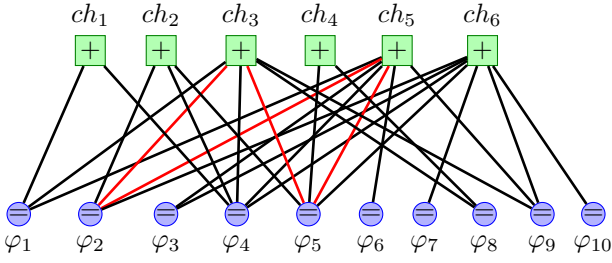
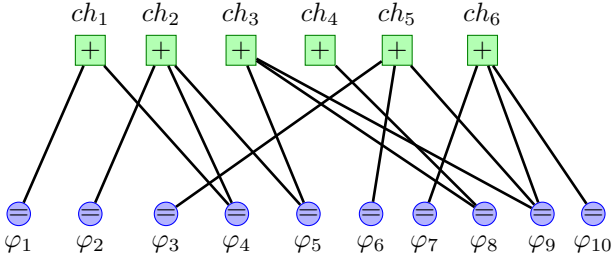
\begin{figure}[t]
    \centering
  \resizebox{0.95\columnwidth}{!}{\begin{tikzpicture}

\tikzset{variable_node_style/.style={circle,draw=blue,fill=blue!30!white, inner sep = 0.1pt}};
\tikzset{check_node_style/.style={regular polygon,regular polygon sides=4,draw=green!50!black,fill=green!30!white, inner sep = 0.1pt}};
\tikzset{edge_style/.style={draw=black,line width=1pt }};

\node [variable_node_style, label=below:{$\varphi_{1}$}] (x1) at (-2.416590,2) {$=$};
\node [variable_node_style, label=below:{$\varphi_{2}$}] (x2) at (-1.467350,2) {$=$};
\node [variable_node_style, label=below:{$\varphi_{3}$}] (x3) at (-0.463346,2) {$=$};
\node [variable_node_style, label=below:{$\varphi_{4}$}] (x4) at (0.476767,2) {$=$};
\node [variable_node_style, label=below:{$\varphi_{5}$}] (x5) at (1.426007,2) {$=$};
\node [variable_node_style, label=below:{$\varphi_{6}$}] (x6) at (2.256592,2) {$=$};
\node [variable_node_style, label=below:{$\varphi_{7}$}] (x7) at (2.968522,2) {$=$};
\node [variable_node_style, label=below:{$\varphi_{8}$}] (x8) at (3.762598,2) {$=$};
\node [variable_node_style, label=below:{$\varphi_{9}$}] (x9) at (4.529291,2) {$=$};
\node [variable_node_style, label=below:{$\varphi_{10}$}] (x10) at (5.204712,2) {$=$};

\node [check_node_style,label=above:{$ch_{1}$}] (c1) at (-1.467350,4.17) {$+$};
\node [check_node_style,label=above:{$ch_{2}$}] (c2) at (-0.527237,4.17) {$+$};
\node [check_node_style,label=above:{$ch_{3}$}] (c3) at (0.531530,4.17) {$+$};
\node [check_node_style,label=above:{$ch_{4}$}] (c4) at (1.581171,4.17) {$+$};
\node [check_node_style,label=above:{$ch_{5}$}] (c5) at (2.603429,4.17) {$+$};
\node [check_node_style,label=above:{$ch_{6}$}] (c6) at (3.735216,4.17) {$+$};

\path [edge_style] (x1) -- (c1);
\path [edge_style] (x2) -- (c2);
\path [edge_style] (x3) -- (c5);
\path [edge_style] (x4) -- (c1);
\path [edge_style] (x4) -- (c2);
\path [edge_style] (x5) -- (c2);
\path [edge_style] (x5) -- (c3);
\path [edge_style] (x6) -- (c5);
\path [edge_style] (x7) -- (c6);
\path [edge_style] (x8) -- (c3);
\path [edge_style] (x8) -- (c4);
\path [edge_style] (x9) -- (c3);
\path [edge_style] (x9) -- (c5);
\path [edge_style] (x9) -- (c6);
\path [edge_style] (x10) -- (c6);
\end{tikzpicture}
  }
    \caption{Tanner graph structure after application of Algorithm~\ref{girth4} for girth-4 cycle removal.}
    \label{Tanner2}
    \vspace{-0.3cm}
\end{figure}
\end{example}

\subsection{Multistage Decoding of Construction-D$'$ Lattices}
The multistage decoding for Construction-D$'$
lattices was first introduced in \cite{15fan}, in which re-encoding by the generator matrix was required during the decoding process. In this algorithm, a binary decoder of the code  $C_{i}$  produces an estimation of the information bits $\widehat{\mathbf{u}}_{i}$'s.
Following this, the re-encoding of $\widehat{\mathbf{u}}_{i}$ through the generator matrix of $C_i$, referred to as $\mathbf{G}_i$, yields the estimated component $\widehat{\mathbf{x}}_{i} = \mathbf{G}_{i} \cdot \widehat{\mathbf{u}}_{i}$. As demonstrated in \cite{silva}, it is feasible to compute the coset in lieu of performing re-encodings within the hypercubical lattice constellations derived from Construction-D$'$.

In this study, the decoding method employed is predicated on the approach detailed in \cite{fan}, which serves as an augmentation of the framework established in \cite{25fan}. It utilizes a sequential cancellation decoding algorithm specifically tailored for Construction-D$'$. The process of re-encoding is executed utilizing the check matrix, thereby allowing for the implementation of non-hypercubical constellations.

Consider a scenario where a lattice point denoted as $\mathbf{x}$ is transmitted and $\mathbf{y} = \mathbf{x} + \mathbf{w}$ represents the vector received from an AWGN channel. In this case, the Gaussian noise vector $\mathbf{w}$ is made up of independent, identically distributed (i.i.d.) Gaussian random variables with variance $\sigma^{2}$ and a mean of zero.

At each level $i$, the decoding process is carried out recursively, for $i = 0, 1, \cdots, a$.  Generally speaking, successive cancellation decoding initializes the $i^{th}$ level of decoding using the outcome from the preceding level $i-1$. The input for each level is a vector $\mathbf{y}_{i} \in [0,1]^{n}$ which will be mapped to $\mathbf{y}_{i}'$, using
\(\mathbf{y}_{i}' = \left| (\mathbf{y}_{i} + \mathbf{1}_{n\times 1}) \hspace{-0.2cm}\pmod{2} - \mathbf{1}_{n\times 1} \right|\),
in which $\mathbf{1}_{n\times 1}$ is the all-one vector of length $n$. Then, we compute the log-likelihood ratio (LLR) vector, for which the $j^{th}$ component is defined as follows:
$$\mathrm{LLR}(j) = \log\left(\frac{\mathrm{Pr}(\widetilde{x}_{i,j} = 0 \mid y_{i,j}')}{\mathrm{Pr}(\widetilde{x}_{i,j} = 1 \mid y_{i,j}')}\right)
	=\frac{1 - 2y_{i,j}'}{2\sigma^{2}},$$
where $x_{i,j}\equiv \widetilde{x}_{i,j} \pmod{2}$ and $x_{i,j}$ is the $j^{th}$ component of $\mathbf{x}_i$.
Afterward, the binary decoding algorithm corresponding to the code
$C_{i}$, denoted by $\text{Dec}_{i}$, for $i = 0, \ldots, a-1$,
is employed that finds the nearest binary codeword
$\widehat{\widetilde{\mathbf{x}}}_{i}$
to $\mathbf{y}_{i}'$
as its output. It should be noted that the objective of lattice decoding is to estimate \(\widehat{\mathbf{x}}_{i}\), while \(\widehat{\widetilde{\mathbf{x}}}_{i}\) serves as the estimate for \(\widetilde{\mathbf{x}}_{i}\).

In the subsequent steps of the decoding process, the estimate \(\widehat{\mathbf{u}}_{i}'\) is first computed using the relation
\begin{equation}\label{uhati}\widehat{\mathbf{u}}_{i}' = \widetilde{H} \odot \widehat{\widetilde{\mathbf{x}}}_{i}.\end{equation}
where \(\odot\) indicates binary multiplication.
 Following this, the estimate \(\widehat{\mathbf{x}}_{i}\) is derived by re-encoding through the check matrix. This re-encoding requires solving the equations
\begin{equation}\label{uhati2}\widetilde{H} \cdot \widehat{\mathbf{x}}_{i} = \widehat{\mathbf{u}}_{i}'.\end{equation}
for \(i = 0, \ldots, a-1\). For the next level of decoding, the estimate \(\widehat{\mathbf{x}}_{i}\) is utilized to prepare the input for the subsequent level which is $\mathbf{y}_{i+1}=\frac{(\mathbf{y}_{i} - \widehat{\mathbf{x}}_{i})}{2}$.
The successive cancellation decoding process concludes when the final component \(\widehat{\mathbf{x}}_{a}\) is obtained. At this last level, the final component of the lattice point is calculated as
\[
\widehat{\mathbf{x}}_{a} = \left\lfloor \mathbf{y}_{a} \right\rceil=\left\lfloor \frac{\mathbf{y}_{a-1} - \widehat{\mathbf{x}}_{a-1}}{2} \right\rceil.
\]
As demonstrated, the estimation of the lattice point \(\mathbf{x}\) in the Construction-D$'$ lattice can be expressed as follows:
\[
\widehat{\mathbf{x}} = \widehat{\mathbf{x}}_{0} + 2\widehat{\mathbf{x}}_{1} + \cdots + 2^{a}\widehat{\mathbf{x}}_{a}.
\]
This process of successive cancellation decoding is detailed in \cite{fan}. It is important to note that the components \(\mathbf{x}_i\) are not necessarily binary.

\subsection{Encoding and Decoding Complexity Analysis}
The decoder combines QC-LDPC belief propagation, LT peeling on the modified Tanner graph, and multistage Construction-D$'$ cancellation. QC-LDPC decoding costs $O(\mathfrak{I}\cdot N)$; LT decoding adds $O(N\log N)$; multistage re-encoding at level $i$ contributes terms scaling with the sparsity of $\widetilde{\mathbf{H}}_i$ in \eqref{uhati}--\eqref{uhati2}.
Since our Raptor parity-check matrix is in approximate lower triangular (ALT) form, encoding and decoding remain linear in the lattice dimension \cite{silva}.
\begin{proposition}[\cite{silva}, Proposition~2 and Proposition~3]\label{decodingprop}
Decoding of the lattice code
\(
\mathcal{C} = \Lambda \cap [0, 2^a)^n,
\)
where \(\Lambda\) is a Construction-D$'$ lattice based on the chain of codes \(C_0 \subset C_1 \subset \cdots \subset C_{a-1}\), can be performed in \(O(a \cdot n)\) operations when each code \(C_i\) admits linear-time decoding and additionally, when encoding of \(\mathcal{C}\) can also be done  within \(O(a \cdot n)\) operations, which holds when the matrices \(\mathbf{H}_i^{p}\) are in an approximate lower triangular (ALT) form, i.e.,
\begin{equation}\mathbf{H}_i^{p} = \begin{bmatrix}\mathbf{B} & \mathbf{L} \\ \mathbf{D} & \mathbf{E}\end{bmatrix},\end{equation}
where \(\mathbf{L} \in \{0,1\}^{(k_i - \mathfrak{g}) \times (k_i - \mathfrak{g})}\) is a lower triangular matrix with ones on its diagonal, and \(\mathfrak{g}\), known as the gap of the ALT form, is of order \(O(1)\).
\end{proposition}
\begin{corollary}[Raptor lattice encoder--decoder complexity]\label{cor:raptor_complexity}
For the proposed $a$-level Construction-D$'$ Raptor lattice with QC-LDPC precode length $N$, LT peeling cost $O(N\log N)$ per stage, and $\mathfrak{I}$ belief-propagation iterations at each level, multistage encoding and decoding require $O(a\mathfrak{I}N + aN\log N + an)$ operations when \eqref{Raptor_parity_check} is in ALT form.
\end{corollary}
\begin{prof}
Summing the per-level QC-LDPC, LT, and re-encoding costs and applying Proposition~\ref{decodingprop} yields the bound.
\end{prof}

The Raptor parity-check matrix in \eqref{Raptor_parity_check} satisfies the ALT structure required by Corollary~\ref{cor:raptor_complexity}.
\begin{remark}
The girth removal technique impacts any quasi-cyclic structure of the underlying codes (at least the QC-LDPC part) in our Raptor lattices. Nonetheless, since our encoding utilizes the ALT-form parity-check matrix of the Raptor code, the overall process remains linear in the lattice dimension, with complexity \(O(a \cdot n)\).
\end{remark}
\section{Secure Cooperative V2X Transmission Protocol}\label{sec_v2x_protocol}
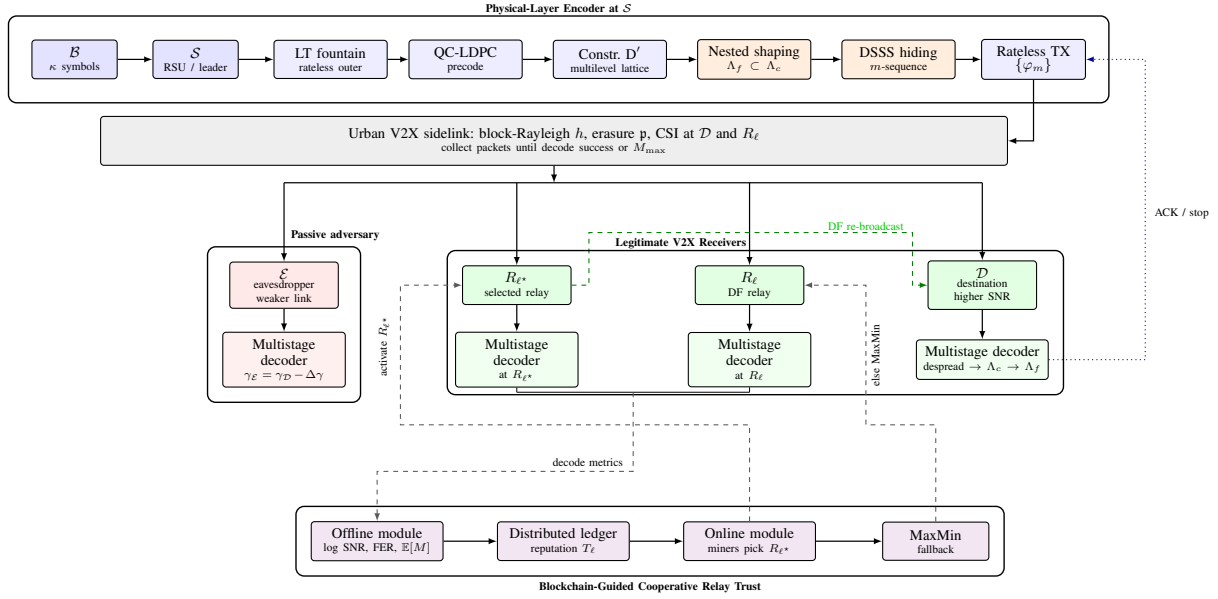
\begin{figure*}[t]
    \centering
    \resizebox{0.88\textwidth}{!}{%
    \begin{tikzpicture}[
        font=\small,
        blk/.style={draw, rounded corners=2pt, align=center, minimum height=0.82cm, inner sep=4pt, line width=0.65pt, text width=2.15cm},
        src/.style={blk, fill=blue!11, text width=1.55cm},
        enc/.style={blk, fill=blue!7},
        sec/.style={blk, fill=orange!13},
        ch/.style={blk, fill=gray!13, text width=19.2cm, minimum height=1.05cm},
        ok/.style={blk, fill=green!11, text width=2.05cm},
        adv/.style={blk, fill=red!9, text width=2.05cm},
        bc/.style={blk, fill=violet!10, text width=2.55cm},
        arr/.style={->, >=latex, line width=0.75pt},
        darr/.style={->, >=latex, line width=0.65pt, dashed, gray!70!black},
        dfarr/.style={->, >=latex, line width=0.65pt, dashed, green!50!black},
        fb/.style={->, >=latex, line width=0.65pt, dotted, blue!60!black},
        lbl/.style={font=\scriptsize, align=center},
        region/.style={draw, rounded corners=5pt, line width=0.85pt, inner sep=14pt}
    ]
        \node[src] (B) at (0,2.3) {$\mathcal{B}$\\[-2pt]\scriptsize $\kappa$ symbols};
        \node[src] (S) at (2.6,2.3) {$\mathcal{S}$\\[-2pt]\scriptsize RSU / leader};
        \node[enc] (lt) at (5.5,2.3) {LT fountain\\[-2pt]\scriptsize rateless outer};
        \node[enc] (qc) at (8.4,2.3) {QC-LDPC\\[-2pt]\scriptsize precode};
        \node[enc] (lat) at (11.5,2.3) {Constr.\ D$'$\\[-2pt]\scriptsize multilevel lattice};
        \node[sec] (nest) at (14.6,2.3) {Nested shaping\\[-2pt]\scriptsize $\Lambda_f \subset \Lambda_c$};
        \node[sec] (dsss) at (17.7,2.3) {DSSS hiding\\[-2pt]\scriptsize $m$-sequence};
        \node[enc, text width=1.95cm] (phi) at (20.6,2.3) {Rateless TX\\[-2pt]$\{\varphi_m\}$};

        \draw[arr] (B.east) -- (S.west);
        \draw[arr] (S.east) -- (lt.west);
        \draw[arr] (lt.east) -- (qc.west);
        \draw[arr] (qc.east) -- (lat.west);
        \draw[arr] (lat.east) -- (nest.west);
        \draw[arr] (nest.east) -- (dsss.west);
        \draw[arr] (dsss.east) -- (phi.west);

        \node[region, fit=(B)(phi), label={[lbl, yshift=-1mm]above:\textbf{Physical-Layer Encoder at} $\mathcal{S}$}] (phybox) {};

        \node[ch] (chan) at (10.3,0.55)
            {Urban V2X sidelink: block-Rayleigh $h$, erasure $\mathfrak{p}$, CSI at $\mathcal{D}$ and $R_\ell$\\[-2pt]
             \scriptsize collect packets until decode success or $M_{\max}$};
        \draw[arr] (phi.south) |- (chan.east);

        \node[adv] (E) at (4.5,-2.55)
            {$\mathcal{E}$\\[-2pt]\scriptsize eavesdropper\\[-2pt]\scriptsize weaker link};
        \node[ok] (Rstar) at (9.5,-2.55) {$R_{\ell^\star}$\\[-2pt]\scriptsize selected relay};
        \node[ok] (Rl) at (14.5,-2.55) {$R_\ell$\\[-2pt]\scriptsize DF relay};
        \node[ok] (D) at (19.5,-2.55)
            {$\mathcal{D}$\\[-2pt]\scriptsize destination\\[-2pt]\scriptsize higher SNR};

        \node[blk, fill=red!6, text width=2.35cm] (decE) at (4.5,-4.15)
            {Multistage decoder\\[-2pt]\scriptsize $\gamma_{\mathcal{E}}\!=\!\gamma_{\mathcal{D}}\!-\!\Delta\gamma$};
        \node[blk, fill=green!6, text width=2.35cm] (decRs) at (9.5,-4.15)
            {Multistage decoder\\[-2pt]\scriptsize at $R_{\ell^\star}$};
        \node[blk, fill=green!6, text width=2.35cm] (decR) at (14.5,-4.15)
            {Multistage decoder\\[-2pt]\scriptsize at $R_\ell$};
        \node[blk, fill=green!6, text width=2.55cm] (decD) at (19.5,-4.15)
            {Multistage decoder\\[-2pt]\scriptsize despread $\rightarrow$ $\Lambda_c$ $\rightarrow$ $\Lambda_f$};

        \draw[arr] (chan.south) -- (10.3,-0.35);
        \draw[-,line width=0.75pt] (4.5,-0.35) -- (19.5,-0.35);
        \draw[arr] (4.5,-0.35) -- (E.north);
        \draw[arr] (9.5,-0.35) -- (Rstar.north);
        \draw[arr] (14.5,-0.35) -- (Rl.north);
        \draw[arr] (19.5,-0.35) -- (D.north);

        \draw[arr] (E) -- (decE);
        \draw[arr] (Rstar) -- (decRs);
        \draw[arr] (Rl) -- (decR);
        \draw[arr] (D) -- (decD);

        \draw[dfarr] (Rstar.east) -- (11.05,-2.55) -- (11.05,-1.42) -- (18.0,-1.42)
            -- (18.0,-2.55) -- (D.west);
        \node[lbl, fill=white, inner sep=1pt,xshift=2.5cm] at (14.5,-1.28) {\textcolor{green!80!black}{DF re-broadcast}};

        \node[region, fit=(Rstar)(decD), inner sep=9pt,
            label={[lbl, yshift=-1mm,xshift=-1.6cm]above:\textbf{Legitimate V2X Receivers}}] (legbox) {};
        \node[region, fit=(E)(decE), inner sep=9pt,
            label={[lbl, yshift=0mm,xshift=1.1cm]above:\textbf{Passive adversary}}] (evebox) {};

        \draw[fb] (decD.east) -- (23.0,-4.15) -- (23.0,2.3) -- (phi.east);
        \node[lbl, anchor=west] at (23.08,-1.0) {ACK / stop};

        \node[bc] (offline) at (6.5,-8.05)
            {Offline module\\[-2pt]\scriptsize log SNR, FER, $\mathbb{E}\lbrack M\rbrack$};
        \node[bc] (ledger) at (10.5,-8.05)
            {Distributed ledger\\[-2pt]\scriptsize reputation $T_\ell$};
        \node[bc] (online) at (14.5,-8.05)
            {Online module\\[-2pt]\scriptsize miners pick $R_{\ell^\star}$};
        \node[bc, text width=2.0cm] (maxmin) at (18.5,-8.05)
            {MaxMin\\[-2pt]\scriptsize fallback};

        \node[region, fit=(offline)(maxmin), inner sep=9pt,
            label={[lbl, yshift=0mm]below:\textbf{Blockchain-Guided Cooperative Relay Trust}}] (bcbox) {};

        \coordinate (metMerge) at (12.0,-4.85);
        \draw (decRs.south) -- (decRs.south |- metMerge);
        \draw (decR.south) -- (decR.south |- metMerge);
        \draw (decRs.south |- metMerge) -- (decR.south |- metMerge);
        \draw[darr] (metMerge) -- (12.0,-6.55) -- (6.5,-6.55) -- (offline.north);
        \node[lbl, anchor=west,yshift=-0.8cm,xshift=-2cm] at (12.15,-5.55) {decode metrics};

        \draw[arr] (offline.east) -- (ledger.west);
        \draw[arr] (ledger.east) -- (online.west);
        \draw[arr] (online.east) -- (maxmin.west);

        \draw[darr] (online.north) -- (14.5,-5.55) -- (7.0,-5.55)
            -- (7.0,-2.55) -- (Rstar.west);
        \node[lbl, rotate=90, anchor=south,xshift=2mm] at (6.88,-4.05) {activate $R_{\ell^\star}$};

        \draw[darr] (maxmin.north) -- (18.5,-5.55) -- (17.0,-5.55)
            -- (17.0,-2.55) -- (Rl.east);
        \node[lbl, rotate=90, anchor=south,yshift=-3mm] at (17.12,-4.05) {else MaxMin};

    \end{tikzpicture}%
    }
    \caption{Proposed cooperative V2X broadcast architecture. At the source $\mathcal{S}$, application block $\mathcal{B}$ is mapped to a rateless Construction-D$'$ Raptor lattice (LT fountain plus QC-LDPC precode), nested-lattice shaped, and direct-sequence spread before transmission over a block-Rayleigh sidelink with erasures. Destination $\mathcal{D}$ and decode-and-forward relays collect rateless packets until multistage lattice decoding succeeds within $M_{\max}$. The passive eavesdropper $\mathcal{E}$ observes the broadcast at reduced SNR. Relay-side decoding statistics are logged offline, stored on a distributed ledger as trust scores $T_\ell$, and used online by miners to activate the best relay $R_{\ell^\star}$, with MaxMin relaying as fallback.}
    \label{fig:v2x_arch}
\end{figure*}

This section specifies how the rateless Raptor lattice codec developed in Sections~\ref{sec_preliminary}--\ref{Sec_decoding} is operated within the cooperative V2X network of Section~\ref{sec_v2x}, integrating nested-lattice PLS, DSSS hiding, and blockchain-assisted DF relay selection \cite{ref202,ref203,ghourab2023}.
Figure~\ref{fig:v2x_arch} summarizes the end-to-end architecture; Section~\ref{sec_simulation} validates the physical-layer codec, fading--erasure delivery performance, and secure broadcast margin under the channel model of Section~\ref{sec_v2x}.
The source $\mathcal{S}$ maps each application block $\mathcal{B}$ to nested-lattice packets with DSSS spreading, emits rateless Construction-D$'$ Raptor lattice packets until $\mathcal{D}$ or an assisting relay acknowledges successful multistage decoding, and relies on coherent despreading together with CSI-aided equalization at legitimate receivers (Fig.~\ref{LFSR}).
The protocol is designed to achieve three objectives: (i)~adaptive sidelink occupancy under time-varying $\gamma_u$ in \eqref{eq:snr_v2x}, (ii)~confidential broadcast delivery when $\gamma_{\mathcal{E}}<\gamma_{\mathcal{D}}$, and (iii)~reliable reception despite random erasures with probability $\mathfrak{p}$ in \eqref{eq:v2x_fading}.
\subsection{Nested-Lattice PLS over V2X Fading Links}
Let \(\mathbf{x}'' \in \Lambda_f \subset \mathbb{R}^{\kappa}\) denote the fine lattice payload from $\mathcal{B}$.
Consider a coarser lattice \(\Lambda_c \subset \Lambda_f\) with generator matrix \(\mathbf{G}_{\Lambda_c} = \mathcal{L} \mathbf{G}_{\Lambda_f}\), where \(\mathcal{L}\) is a positive nesting index.
We select a random dithering vector \(\mathbf{r} \in \Lambda_c\) and define
\[
\mathbf{x}' = \left[ \mathbf{x}'' + \mathbf{r} \right]\,\, (\bmod\,\,\Lambda_c).
\]
Since we have performed nested lattice shaping \((\bmod\, \Lambda_c)\), the vector \(\mathbf{x}'\) satisfies the power constraint \(P\).
To prevent fine-lattice recovery under eavesdropper channel conditions, denote by \(\sigma_f^2\) the noise variance at which fine-lattice decoding achieves the target FER (e.g., $10^{-4}$), and transmit with coarse noise variance \(\sigma_c^2 = \mathcal{L}^2 \sigma_f^2\).
With \(\mathbf{G}_{\Lambda_c}=\mathcal{L}\mathbf{G}_{\Lambda_f}\), the volume-to-noise ratio (VNR) is unchanged across nesting levels because \(\det(\mathbf{G}_{\Lambda_c})=\mathcal{L}^{\kappa}\det(\mathbf{G}_{\Lambda_f})\) cancels \(\mathcal{L}^2\) in \(\mathrm{VNR}(\Lambda)=\det(\mathbf{G}_{\Lambda})^{2/\kappa}/(2\pi e\,\sigma^2)\).
Secure V2X broadcast therefore requires a persistent channel-quality advantage at $\mathcal{D}$ over $\mathcal{E}$, as enforced by path loss, beamforming, or the SNR deficit $\Delta\gamma$ in Section~\ref{sec_v2x}; the formal secrecy analysis is completed after the despreading model below.
Let \(\mathbf{s}[\tau] \in \{-1, +1\}^{\mathfrak{s}}\) be a random sequence with \(\pm 1\) components of length \(\mathfrak{s}\) at time interval $\tau$, where $\tau$ counts the number of transmitted symbols to the channel.
We assume that \(\mathbf{s}[\tau]\) is derived from the \(\tau^{th}\) state of a binary linear feedback shift register (LFSR) generating an \(m\)-sequence, with the mapping \(0 \to -1\) and \(1 \to +1\) \cite{DSSS}.

A maximum-length sequence (MLS) or \(m\)-sequence is a binary pseudorandom sequence produced by an LFSR of length \(\mathfrak{s}\) whose period is maximal \((2^{\mathfrak{s}} - 1)\).
Indeed, an \(m\)-sequence \(\{s_i\}_{i \ge 0}\) is generated by an \(\mathfrak{s}\)-stage LFSR over \(\mathbb{F}_2\) with connection polynomial \cite{DSSS}
\begin{equation}
s(x) = 1 + c_1 x + c_2 x^2 + \cdots + c_{\mathfrak{s}-1} x^{\mathfrak{s}-1} + x^{\mathfrak{s}},
\end{equation}
where \(s(x)\) is primitive over \(\mathbb{F}_2\), and the sequence satisfies
\begin{equation*}
\begin{aligned}
s_{i + \mathfrak{s}}
&= c_1\cdot s_{i + \mathfrak{s}-1}
\oplus c_2 \cdot s_{i + \mathfrak{s}-2}\\
&\quad\oplus \cdots \oplus c_{\mathfrak{s}-1}\cdot s_{i+1} \oplus s_i.
\end{aligned}
\end{equation*}
If the LFSR is initialized with a nonzero state, the resulting sequence has period \(2^{\mathfrak{s}} - 1\), and is called an \(m\)-sequence \cite{DSSS}.

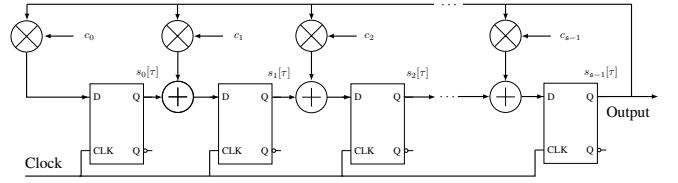
\begin{figure}[tb]
    \centering
    \resizebox{0.98\columnwidth}{!}{
\begin{tikzpicture}
	\node[latch, dot on notQ] at (2.88, 6.24){};
	\node[latch, dot on notQ] at (6.92, 6.24){};
	\node[latch, dot on notQ] at (11.08, 6.24){};
	\node[latch, dot on notQ] at (17.147, 6.259){};
	\node[adder] at (4.8, 7.09){};
	\node[adder] at (4.8, 7.09){};
	\node[adder] at (9, 7.08){};
	\node[adder] at (15.1, 7.09){};
	\draw[line width=1pt, -latex] (7.8, 7.08) |- (8.51, 7.08);
	\draw[line width=1pt] (19.059, 7.1) -- (19.059, 10) -- (0.059, 10);
	\node[mixer] at (0.046, 8.99){};
	\node[mixer] at (4.79, 9){};
	\draw[line width=1pt, -latex] (18, 7.1) |- (19.9, 7.1);
	\node[mixer] at (9, 9.01){};
	\node[mixer] at (15.1, 9.01){};
	\draw[line width=1pt, -latex] (15.6, 7.1) -- (16.3, 7.1);
	\draw[line width=1pt, -latex] (15.1, 10) -| (15.1, 9.5);
	\draw[line width=1pt, -latex] (15.1, 8.5) -| (15.1, 7.58);
	\draw[line width=1pt, -latex] (0.046, 8.481) |- (2, 7.081);
	\draw[line width=1pt, -latex] (11.9, 7.086) -- (12.9, 7.086);
	\draw[line width=1pt, -latex] (13.6, 7.1) -- (14.6, 7.1);
	\draw[line width=1pt, -latex] (9.49, 7.08) |- (10.233, 7.08);
	\draw[line width=1pt, -latex] (5.29, 7.09) |- (6.099, 7.081);
	\draw[line width=1pt, -latex] (9, 10) -| (9, 9.5);
	\draw[line width=1pt, -latex] (9, 8.52) -| (9, 7.6);
	\draw[line width=1pt, -latex] (4.79, 9.99) -| (4.79, 9.49);
	\draw[line width=1pt, -latex] (4.8, 8.5) -| (4.8, 7.58);
	\draw[line width=1pt, -latex] (0.059, 10) -| (0.059, 9.5);
	\draw[line width=1pt, -latex] (6.22, 9) |- (5.3, 9);
	\draw[line width=1pt, -latex] (1.52, 9) |- (0.6, 9);
	\draw[line width=1pt, -latex] (10.42, 9) |- (9.5, 9);
	\draw[line width=1pt, -latex] (16.52, 9) |- (15.6, 9);
	\draw[line width=1pt] (0, 4.6) -- (16, 4.6) -| (16.027, 5.419);
	\draw[line width=1pt] (1.76, 5.4) |- (1.7, 4.6);
	\draw[line width=1pt] (5.8, 5.4) -| (5.8, 4.6);
	\draw[line width=1pt] (9.96, 5.4) |- (10, 4.6);
	\node[shape=rectangle, minimum width=0.9cm, minimum height=0.6cm] at (13.25, 7.1){} node[anchor=north west, align=left, text width=0.456cm, inner sep=6.3pt] at (12.8, 7.4){\large $\cdots$};
	\node[shape=rectangle, minimum width=0.865cm, minimum height=0.565cm] at (2.1, 9){} node[anchor=north west, align=left, text width=0.477cm, inner sep=6pt] at (1.65, 9.3){$c_0$};
	\node[shape=rectangle, minimum width=0.865cm, minimum height=0.565cm] at (6.8, 9){} node[anchor=north west, align=left, text width=0.477cm, inner sep=6pt] at (6.35, 9.3){$c_1$};
	\node[shape=rectangle, minimum width=0.865cm, minimum height=0.565cm] at (10.9, 9){} node[anchor=north west, align=left, text width=0.477cm, inner sep=6pt] at (10.45, 9.3){$c_2$};
	\node[shape=rectangle, minimum width=0.865cm, minimum height=0.565cm] at (17.05, 9){} node[anchor=north west, align=left, text width=0.477cm, inner sep=6pt] at (16.6, 9.3){$c_{\mathfrak{s}-1}$};
	\node[shape=rectangle, minimum width=0.865cm, minimum height=0.565cm] at (3.727, 7.953){} node[anchor=north west, align=left, text width=0.477cm, inner sep=6pt] at (3.277, 8.253){$s_{0}[\tau]$};
	\node[shape=rectangle, minimum width=0.865cm, minimum height=0.565cm] at (7.861, 7.921){} node[anchor=north west, align=left, text width=0.477cm, inner sep=6pt] at (7.411, 8.221){$s_{1}[\tau]$};
	\node[shape=rectangle, minimum width=0.865cm, minimum height=0.565cm] at (12.215, 7.921){} node[anchor=north west, align=left, text width=0.477cm, inner sep=6pt] at (11.765, 8.221){$s_{2}[\tau]$};
	\node[shape=rectangle, fill={rgb,255:red,255;green,255;blue,255}, minimum width=0.9cm, minimum height=0.6cm] at (13.3, 10.007){} node[anchor=north west, align=left, text width=0.456cm, inner sep=6.3pt] at (12.85, 10.307){\large $\cdots$};
	\node[shape=rectangle, minimum width=1.615cm, minimum height=0.565cm] at (18.875, 6.7){} node[anchor=north west, align=left, text width=1.227cm, inner sep=6pt] at (18.05, 7){\Large Output};
	\node[shape=rectangle, minimum width=1.486cm, minimum height=0.565cm] at (0.54, 5.1){} node[anchor=north west, align=left, text width=1.097cm, inner sep=6pt] at (-0.221, 5.4){\Large Clock};
	\node[shape=rectangle, minimum width=0.865cm, minimum height=0.565cm] at (17.807, 7.956){} node[anchor=north west, align=left, text width=0.477cm, inner sep=6pt] at (17.357, 8.256){$s_{\mathfrak{s}-1}[\tau]$};
	\draw[line width=1pt, -latex] (3.71, 7.082) |- (4.31, 7.082);
\end{tikzpicture}
}
    \caption{Implementation of an $m$-sequence generator using an LFSR.}
    \label{LFSR}
    \vspace{-0.6cm}
\end{figure}

Maximum-length sequences provide low cross-correlation spreading sequences \cite{golomb1967shift,DSSS}.
For V2X broadcast, the balance and run properties of an $m$-sequence support reliable despreading at $\mathcal{D}$ while the processing gain and nested shaping increase the effective acquisition barrier for $\mathcal{E}$ when $\gamma_{\mathcal{E}}<\gamma_{\mathcal{D}}$.
To ensure unpredictable state transitions, we use the internal-feedback LFSR implementation in Fig.~\ref{LFSR}.

Changing the LFSR initial state produces a distinct $m$-sequence; with $\mathfrak{s}=64$, the key space exceeds $2^{120}$, comparable to 128-bit symmetric security \cite{DSSS}.

Now, let \(n = \kappa \cdot \mathfrak{s}\). In the initial communication round, the lattice vector \(\mathbf{x}' = (x'_1, \ldots, x'_{\kappa})\) is expanded to \(\mathbf{x} = (x_1, \ldots, x_n) = \mathbf{x}' \circ \mathbf{s}[[0:\kappa]] = \big(x'_1 \cdot \mathbf{s}[1], \ldots, x'_{\kappa} \cdot \mathbf{s}[\kappa]\big)\), where \(\circ\) denotes the Hadamard or entry-wise  product, and \(\mathbf{s}[[\tau_1:\tau_2]] = \big(\mathbf{s}[\tau_1+1], \ldots, \mathbf{s}[\tau_2]\big)\) for \(\tau_1 < \tau_2\). In the \(\tau^{th}\) round, $\mathbf{x}'$ will be expanded to
\begin{equation*}
\begin{aligned}
\mathbf{x}&=\mathbf{x}'\circ\mathbf{s}[[(\tau-1)\kappa:\tau\kappa]]\\
&=\big(x'_1\cdot\mathbf{s}[(\tau-1)\kappa+1],\ldots,x'_{\kappa}\cdot\mathbf{s}[\tau\kappa]\big).
\end{aligned}
\end{equation*}
To meet the power constraints, we must have \(\frac{1}{n} \|\mathbf{x}\|^2 \leq P\), which is satisfied because
\begin{equation*}
\begin{aligned}
\frac{1}{n}\|\mathbf{x}\|^2
&=\frac{\mathfrak{s}|x_1'|^2+\cdots+\mathfrak{s}|x_{\kappa}'|^2}{\kappa\mathfrak{s}}\\
&=\frac{1}{\kappa}\|\mathbf{x}'\|^2\le P.
\end{aligned}
\end{equation*}
Two scenarios can be considered: (i)~keep symbol duration and increase bandwidth by $\mathfrak{s}$, spreading the signal as in DSSS; or (ii)~keep bandwidth and lengthen the frame by $\mathfrak{s}$.

\subsubsection{V2X Fading--Erasure Channel Capacity}
Under the V2X model of Section~\ref{sec_v2x}, each lattice packet experiences block fading and erasures as in \eqref{eq:v2x_fading}.
The ergodic capacity of the legitimate link conditioned on successful reception is \cite{Verdu,HassanIT}
\begin{equation}\label{capacity_v2x}
    \bar{C}_{\mathcal{D}}=(1-\mathfrak{p})\,\mathbb{E}\!\left[\log_2\!\left(1+\gamma_{\mathcal{D}}\right)\right],
\end{equation}
where the expectation is over $|h_{\mathcal{D}}|^2$.
Nested Construction-D$'$ Raptor lattice codes operate near this limit when the number of collected rateless packets adapts to the instantaneous $\gamma_{\mathcal{D}}$ and erasure pattern \cite{Erasure,silva}.
For the eavesdropper, replace $\gamma_{\mathcal{D}}$ with the weaker $\gamma_{\mathcal{E}}$ in \eqref{capacity_v2x} to obtain $\bar{C}_{\mathcal{E}}$; the secure broadcast margin is $\big[\bar{C}_{\mathcal{D}}-\bar{C}_{\mathcal{E}}\big]^+$ after DSSS and nested shaping.
The codewords satisfy $\frac{1}{n}\|\mathbf{x}\|^2\le P$ with $\mathbf{x}$ drawn from the nested codebook $\Lambda_f/\Lambda_c$ after spreading; the erasure-limited AWGN capacity is \cite{Verdu,Erasure}
\begin{equation}\label{capacity}
    C=\frac{1-\mathfrak{p}}{2}\log_2\left(1+\frac{P}{\sigma_c^2}\right).
\end{equation}
Our multilevel binary Raptor Construction-D$'$ implementation targets the same limit as nested Construction-A lattice codes \cite{Erasure}.
\subsubsection{Decoding Steps at Legitimate Receiver}
At the receiver, we perform synchronization by utilizing the correlation features of $m$-sequences, which is very similar to the synchronization process in DSSS schemes \cite{DSSS}. Here, we consider the baseband model, and our task is to despread the received vector $\mathbf{y} = (y_1, \ldots, y_n)$. During the $\tau^{th}$ communication interval, we multiply $\mathbf{y}$ by  its corresponding spreading sequence:
\begin{equation*}
\begin{aligned}
\mathbf{y}'&=\mathbb{E}_{\mathfrak{s}}\big[\mathbf{y}\circ\mathbf{s}[[(\tau-1)\kappa:\tau\kappa]]\big]\\
&=\Big(\mathbb{E}\big[y_1 s^{[(\tau-1)\kappa+1]}_{1},\ldots,y_{\mathfrak{s}}s^{[(\tau-1)\kappa+1]}_{\mathfrak{s}}\big],\\
&\qquad\ldots,\mathbb{E}\big[y_{(\kappa-1)\mathfrak{s}+1}s^{[\tau\kappa]}_{1},\ldots,y_{\kappa\mathfrak{s}}s^{[\tau\kappa]}_{\mathfrak{s}}\big]\Big).
\end{aligned}
\end{equation*}
where $\mathbb{E}$ denotes the expectation, and $\mathbb{E}_{\mathfrak{s}}$ represents the operator that replaces every $\mathfrak{s}$ symbols with their average.   Now, we expand the first block of size $\mathfrak{s}$. If a symbol is erased, we replace it with zero. As long as there are enough un-erased symbols remaining within a block, we can successfully complete the despreading. Let $\mathfrak{s}_i$ denote the number of un-erased symbols in block $i$. Without loss of generality, let $y_i = x_i + z_i$ for $i = 1, \ldots, \mathfrak{s}_1$, and $y_i = \xi$ for $i = \mathfrak{s}_1 + 1, \ldots, \mathfrak{s}$. Since for $i = 1, \ldots, \mathfrak{s}_1$, we have $x_i = x_1's^{[(\tau - 1) \cdot \kappa + 1]}_{i}$, it follows that
\begin{equation*}
\begin{aligned}
&\mathbb{E}\Big[y_1 s^{[(\tau-1)\kappa+1]}_{1},\ldots,y_{\mathfrak{s}}s^{[(\tau-1)\kappa+1]}_{\mathfrak{s}}\Big]\\
&\quad=\frac{1}{\mathfrak{s}_1}\sum_{i=1}^{\mathfrak{s}_1}y_i s^{[(\tau-1)\kappa+1]}_{i}\\
&\quad=\frac{1}{\mathfrak{s}_1}\sum_{i=1}^{\mathfrak{s}_1}\big(x_1's^{[(\tau-1)\kappa+1]}_{i}+z_i\big)s^{[(\tau-1)\kappa+1]}_{i}\\
&\quad=x_1'+z_1',
\end{aligned}
\end{equation*}
where $z_1' = \frac{\sum_{i=1}^{\mathfrak{s}_1} z_i s^{[(\tau-1)\cdot\kappa+1]}_{i}}{\mathfrak{s}_1} \sim \mathcal{N}\left(0, \frac{\mathfrak{s}_1\sigma_c^2}{\mathfrak{s}_1^2}\right)$. The remaining blocks can be despread similarly, resulting in $\mathbf{x}_{DS}' = \mathbf{x}' + (z_1', \ldots, z_{\kappa}')$, where $z_i' \sim \mathcal{N}\left(0, \frac{\mathcal{L}^2 \sigma_f^2}{\mathfrak{s}_i}\right)$.

By considering the erasure probability $\mathfrak{p}$, we can select $\mathcal{L} $ and $\mathfrak{s}$ such that $\frac{\mathcal{L} ^2 \sigma_f^2}{\mathfrak{s}_i} \approx \sigma_f^2$ for $i = 1, \ldots, \kappa$, making coarse lattice and fine lattice decoding feasible. Since on average \(\mathfrak{s}_i \approx (1 - \mathfrak{p}) \mathfrak{s}\), we can assume \(\mathcal{L}^2 \approx (1 - \mathfrak{p}) \mathfrak{s}\). This implies that when \(\mathfrak{p} \approx 0\), and based on our previous assumptions, \(\mathcal{L} \approx \sqrt{\mathfrak{s}} \approx 8\).

Given that $\mathbf{x}' = \mathbf{x}'' + \mathbf{r} - \mathbf{r}_S$, where $\mathbf{r}_S \in \Lambda_c$ arises from the nested lattice shaping, providing $\mathbf{x}_{DS}'$ to the decoder of the coarse lattice $\Lambda_c$ yields $\mathbf{r}'' = \hat{\mathbf{r}} - \hat{\mathbf{r}}_S$. By subtracting $\mathbf{r}''$ from $\mathbf{x}_{DS}'$ and supplying the resulting vector to the fine lattice $\Lambda_f$ decoder, we can obtain the estimation of $\mathbf{x}''$. After successful decoding, the value of $\tau$ will be incremented by one in both the transmitter and receiver. To enhance the decoding probability, we can also incorporate scrambling and descrambling at the input and output of the channel to distribute the erasures across different blocks.

\subsubsection{Secrecy Analysis of Rateless Nested-Lattice Broadcast}\label{sec:secrecy_analysis}
We characterize the secure broadcast margin under the full protocol model: block-Rayleigh fading \eqref{eq:snr_v2x}, erasures \eqref{eq:v2x_fading}, nested shaping with index $\mathcal{L}$, DSSS with factor $\mathfrak{s}$, and rateless transmission of up to $M_{\max}$ packets per session.
For packet index $m$ and receiver $u\in\{\mathcal{D},\mathcal{E}\}$, let $\gamma_{u,m}$ denote the post-equalization fine-lattice SNR after despreading.
At the destination, $\gamma_{\mathcal{D},m}=P|h_{\mathcal{D},m}|^2/\sigma_f^2$, while the passive eavesdropper observes the same broadcast index with coupled SNR
\begin{equation}\label{eq:gamma_E}
\begin{aligned}
\gamma_{\mathcal{E},m}&=10^{-\Delta\gamma_{\mathrm{tot}}/10}\,\gamma_{\mathcal{D},m},\\
\Delta\gamma_{\mathrm{tot}}&=\Delta\gamma+\Delta\gamma_{\mathrm{DSSS}}.
\end{aligned}
\end{equation}
where $\Delta\gamma$ is the average path-loss/beamforming deficit in Section~\ref{sec_v2x} and $\Delta\gamma_{\mathrm{DSSS}}$ captures despreading and acquisition loss at $\mathcal{E}$.
For a non-erased packet, let $q^f(\gamma)=\Pr\{\mathrm{FER}>\varepsilon\,|\,\mathrm{VNR}(\gamma)\}$ denote the fine-lattice decode-failure probability; we assume $q^f(\gamma)$ is non-increasing in $\gamma$, as verified by the AWGN $\mathrm{FER}(\mathrm{VNR})$ curve in Fig.~\ref{fig:101}.

\begin{proposition}[Achievable nested-lattice secrecy rate]\label{prop:secrecy_rate}
Under $\mathcal{L}^2=(1-\mathfrak{p})\mathfrak{s}$ and $\sigma_c^2=\mathcal{L}^2\sigma_f^2$, the erasure channel \eqref{eq:v2x_fading} with fine-lattice SNR $\gamma_f=P/\sigma_f^2$ at $\mathcal{D}$ supports
\begin{equation}\label{eq:Rs_expand}
\begin{aligned}
R_s&=(1-\mathfrak{p})\,\mathfrak{s}W\Big[
\log_2\!\big(1+\gamma_f\big)\\
&\qquad-\log_2\!\Big(1+\frac{\gamma_f}{(1-\mathfrak{p})\mathfrak{s}}\Big)
\Big]^{+}
\end{aligned}
\end{equation}
under bandwidth expansion by $\mathfrak{s}$, and the same bracket multiplied by $W$ under fixed bandwidth.
Moreover, $R_s>0$ whenever $(1-\mathfrak{p})\mathfrak{s}>1$.
\end{proposition}
\begin{prof}
After despreading, each non-erased symbol at $\mathcal{D}$ carries at most $(1-\mathfrak{p})\log_2(1+\gamma_f)$ bits by \eqref{capacity}.
Without $\mathbf{r}$ and the spread sequence, $\mathcal{E}$ resolves only coarse cosets at effective SNR $\gamma_f/((1-\mathfrak{p})\mathfrak{s})$ because $\det(\mathbf{G}_{\Lambda_c})=\mathcal{L}^{\kappa}\det(\mathbf{G}_{\Lambda_f})$ and $\sigma_c^2=\mathcal{L}^2\sigma_f^2$.
Subtracting the two bounds yields the stated $R_s$; positivity requires $(1-\mathfrak{p})\mathfrak{s}>1$.
\end{prof}

\begin{theorem}[Secure rateless broadcast over the V2X wiretap channel]\label{thm:secure_rateless}
Consider a session with rateless packets $m=1,\ldots,M_{\max}$, events $E_{u,m}$ (packet~$m$ not erased and fine-lattice decoding succeeds at $u$), and $\mathcal{S}_{u,M}=\bigcup_{m=1}^{M}E_{u,m}$.
With independent block-fading gains $\{|h_{\mathcal{D},m}|^2\}$, a common erasure realization, and SNR coupling \eqref{eq:gamma_E}, (i)~$\Pr\{\mathcal{S}_{\mathcal{D},M}\}$ is non-decreasing in $M$ and $\mathcal{P}_{\mathrm{out}}=\Pr\{\mathcal{S}_{\mathcal{D},M_{\max}}^c\}$ (Theorem~\ref{thm:rateless_mono}); (ii)~$\Pr\{E_{\mathcal{E},m}\}\le \Pr\{E_{\mathcal{D},m}\}$ for every $m$; and (iii)~$\Pr\{\mathcal{S}_{\mathcal{E},M}\}\le \Pr\{\mathcal{S}_{\mathcal{D},M}\}$ for every $M\le M_{\max}$.
\end{theorem}
\begin{prof}
For~(ii), $\Pr\{E_{u,m}\}=(1-\mathfrak{p})[1-q^f(\gamma_{u,m})]$ and \eqref{eq:gamma_E} imply the ordering.
For~(iii), $\Pr\{\mathcal{S}_{u,M}^c\}=\prod_{m=1}^{M}\big(1-(1-\mathfrak{p})[1-q^f(\gamma_{u,m})]\big)$; each factor is larger for $\mathcal{E}$ than for $\mathcal{D}$.
\end{prof}

\subsection{Blockchain-Guided DF Relay Selection in Vehicular Networks}
Building upon the nested lattice PLS design and the trustworthy relaying paradigm of \cite{ref202,ref203,ghourab2023}, road-side infrastructure periodically logs sidelink SNR, FER, and decoding-completion overhead observed at each candidate DF relay during offline monitoring intervals.
These physical-layer statistics are committed to a distributed ledger as reputation records $T_\ell$, which miners update from historical decoding success and average air-interface occupancy $\mathbb{E}\lbrack M\rbrack$.
When $\mathcal{S}$ must deliver $\mathcal{B}$ to $\mathcal{D}$, the online control plane activates the highest-ranked relay $R_{\ell^\star}=\arg\max_{\ell}T_\ell$ for decode-and-forward retransmission, and reverts to MaxMin relay selection whenever no trustworthy record is available \cite{ref203}.
During each broadcast session, candidate relays attempt lattice decoding and report their outcomes; the resulting statistics refine the ranking used for subsequent $\mathcal{S}\!\to\!R_\ell\!\to\!\mathcal{D}$ cooperation.
Rateless reception is particularly compatible with reputation-based relaying because partial packet collection can still support successful decode-and-forward recovery, thereby reducing spurious negative reputation events caused by transient fading dips.

\section{V2X Performance Evaluation}\label{sec_simulation}
This section evaluates the proposed framework in two complementary tiers.
The first tier benchmarks the underlying Raptor lattice codec on an AWGN reference channel at matched block lengths against LDLC and generalized Construction-D$'$ LDPC lattices.
The second tier conducts Monte Carlo evaluation under the block-Rayleigh fading and erasure model of Section~\ref{sec_v2x} with the parameters in Table~\ref{tab:v2x_params}, focusing on the $n=995$ operating point validated in Fig.~\ref{fig:101}.
Fading and PLS curves employ a rateless-aware link-level abstraction: after coherent equalization, each received packet succeeds or fails according to the AWGN $\mathrm{FER}(\mathrm{VNR})$ curves in Fig.~\ref{fig:101}, with Raptor-specific fountain accumulation \eqref{eq:raptor_accum} and fixed-length H-ARQ baselines as described below.
Matched-length LDLC and LDPC lattices are benchmarked under identical block-Rayleigh and erasure realizations; blockchain-assisted relay selection is architecturally specified in Section~\ref{sec_v2x_protocol} but not quantitatively assessed here (Section~\ref{sec:eval_scope}).

\subsection{AWGN Codec Benchmark}
\begin{figure*}[tb]
\centering
\subfloat[SER on the AWGN reference channel ($n\approx 1000$).]{\includegraphics[clip, trim=30 10 50 50, width=0.48\textwidth]{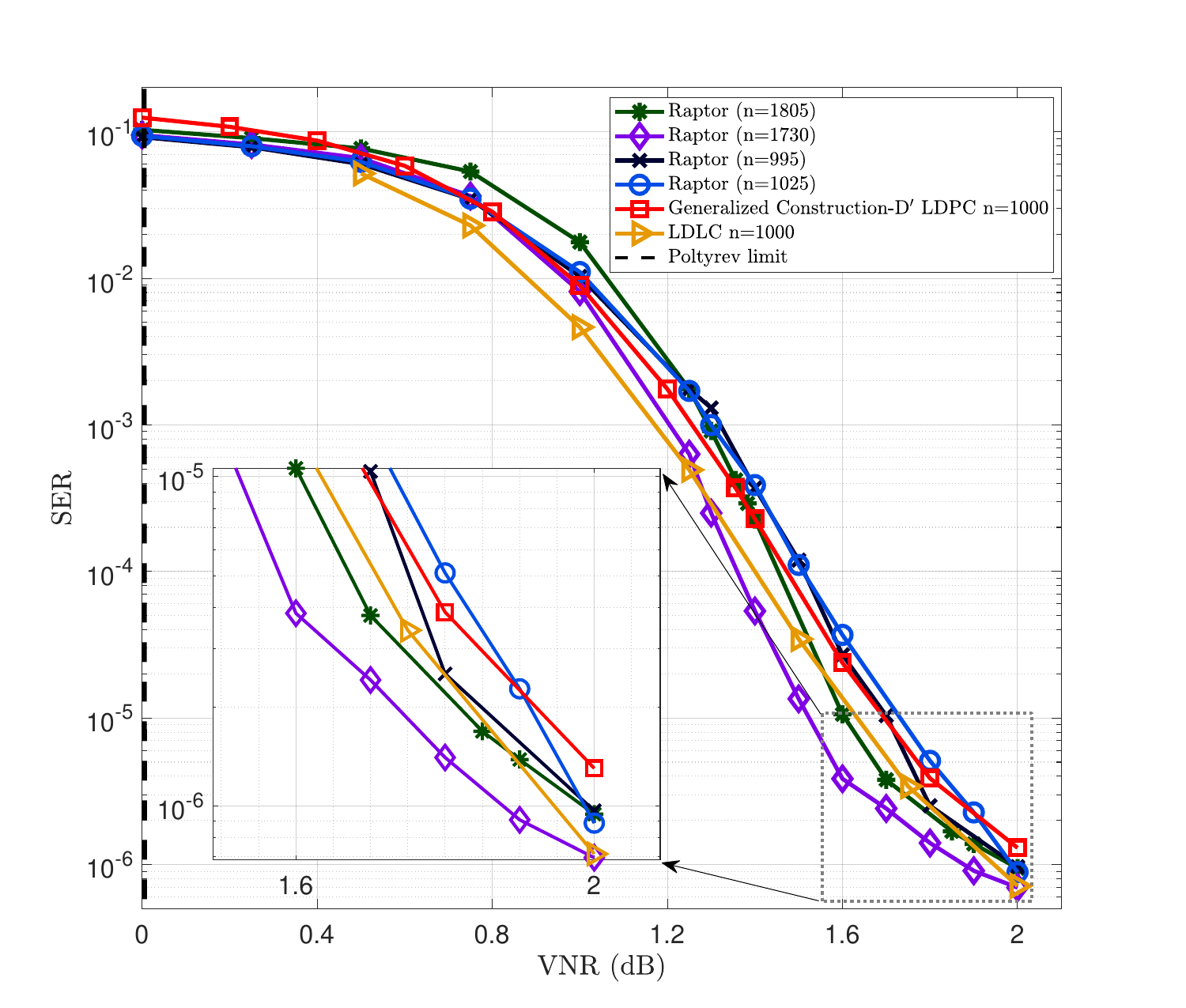}\label{fig:1001}}
\hfill
\subfloat[FER vs.\ LDA, LDPC, LDLC, and Polar.]{\includegraphics[clip, trim=30 10 50 50, width=0.48\textwidth]{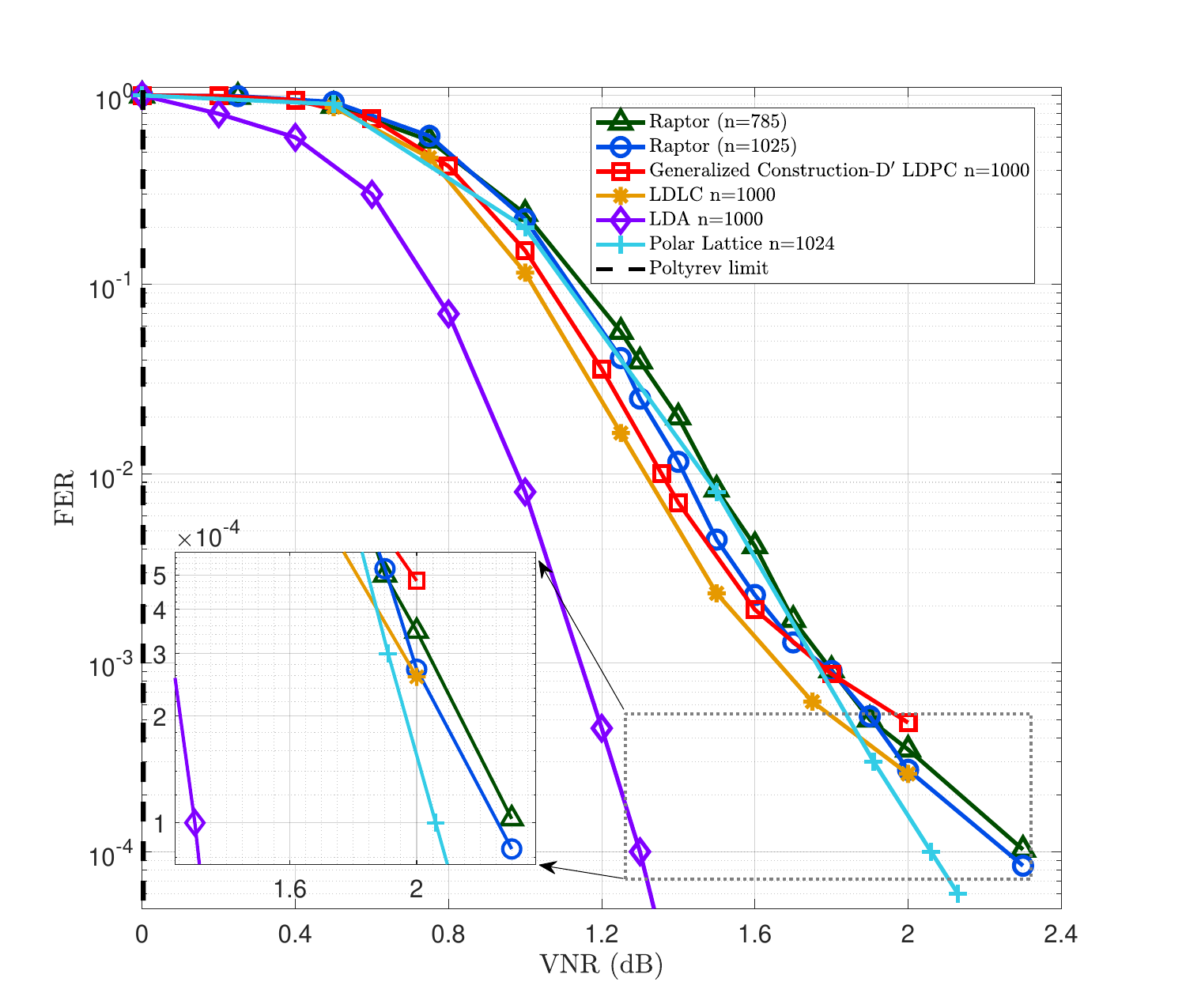}\label{fig:101}}
\caption{AWGN reference-channel benchmarks: (a) SER; (b) FER.}
\end{figure*}

The first evaluation tier characterizes the underlying Raptor lattice codec on an AWGN reference channel with $|h|^2=1$ and $\mathfrak{p}=0$, which models short-range line-of-sight V2X links in which fading and erasures are negligible over the coherence interval of interest.
An infinite sequence of encoded symbols, \(\varphi_1, \varphi_2, \varphi_3, \ldots\), is generated by the rateless fountain encoder; decoding trials terminate once a prescribed number of frame error events has been observed over finite measurement windows.
We evaluate Raptor lattices with lengths \(n = 785, 995, 1025, 1730,\) and \(1805\).
Comparisons at \(n\approx 1000\) employ LDLC and generalized Construction-D$'$ LDPC lattices of length \(1000\) \cite{silva,29silva}; the \(n=13999\) design illustrates asymptotic scaling behavior and is excluded from the matched-length comparison.
Figures~\ref{fig:1001} and~\ref{fig:101} report symbol error rate (SER) and frame error rate (FER) versus volume-to-noise ratio (VNR).

To construct LT codes, following the heuristic of \cite{2019}, we set \(D = 12\) and \(\alpha = 0.27\). Raptor codes are formed by concatenating these LT codes with QC-LDPC codes, and their parameters are reported in Table~\ref{tab:lattice_parameters}. For lattice construction, we configure \(a = 3\). The number of iterations is set to $\mathcal{T}=50$.
\setlength{\tabcolsep}{3pt} 
\renewcommand{\arraystretch}{0.9} 
\begin{table}[h]
    \centering
    \caption{Summary of parameters and performance results for Raptor, LDLC, and LDPC lattices.}
    \resizebox{\columnwidth}{!}{%
    \begin{tabular}{@{}cccc@{}}
        \toprule
        \textbf{Lattice} & $(n,N,K,w_c,w_r,\vartheta)$ & \textbf{SER$= 10^{-5}$} & \textbf{FER$= 10^{-4}$} \\
        \midrule
        Raptor & $(785,540,450,2,12,46)$ & 1.80 dB & 2.30 dB \\
        Raptor & $(995,684,570,2,12,57)$ & 1.70 dB & 2.20 dB\\
        Raptor & $(1025,708,590,2,12,57)$  & 1.73 dB & 2.25 dB \\
        Raptor & $(1730,1188,990,2,12,99)$  & 1.52 dB & 2.38 dB \\
        Raptor & $(1805,1190,935,14,3,85)$  & 1.60 dB & 2.55 dB \\
        Raptor & $(13999, 9600,8000,2,12,800)$ & 1.25 dB & 3.00 dB
                          \\
        LDLC \cite{29silva} & $n=1000$ & 1.63 dB & 2.27 dB \\
        LDPC \cite{silva} &  $(n,R_0,R_1) = (1000,0.5, 0.978)$ & 1.70 dB & 2.52 dB \\
        LDA \cite{Pietro2018} & $n=1000$ & 1.33 dB & 1.30 dB \\
        Polar \cite{Polarm} & $n=1024$ & N/A & 2.06 dB \\
        \bottomrule
    \end{tabular}%
    }
    \label{tab:lattice_parameters}
    \vspace{-0.2cm}
\end{table}
In Table~\ref{tab:lattice_parameters}, SER and FER benchmarks at $\mathrm{SER}=10^{-5}$ and $\mathrm{FER}=10^{-4}$ favor different candidates: at $n=995$, the proposed codec improves FER by $0.07$~dB and $0.32$~dB over matched-length LDLC and LDPC lattices, which is the operating point used below, while LDA and polar designs remain stronger in absolute FER.
Regarding complexity, QC-LDPC precodes and systematic Raptor encoding yield lower encoder/decoder cost than the constructions in \cite{silva,29silva}.

\subsection{Block-Rayleigh Fading and Erasure Performance}
\begin{figure*}[tb]
\centering
\subfloat[Outage probability and mean packet count at $\mathfrak{p}=0.15$.]{\includegraphics[width=0.48\textwidth]{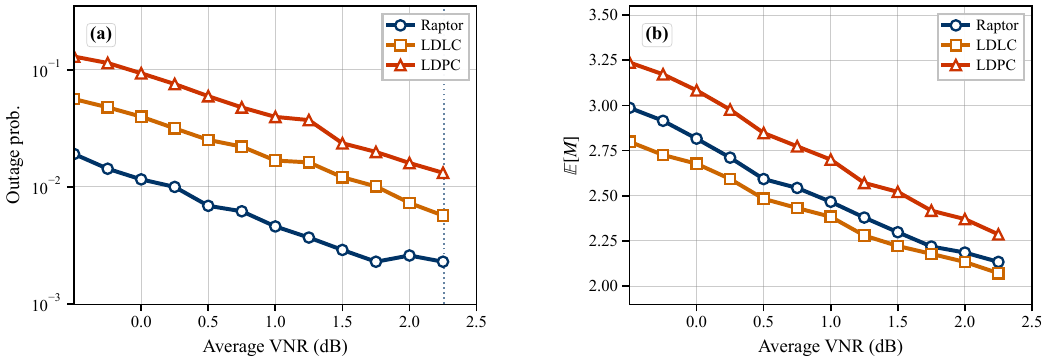}\label{fig:v2x_fading}}
\hfill
\subfloat[Raptor delivery heatmaps over VNR and $\mathfrak{p}$.]{\includegraphics[width=0.48\textwidth]{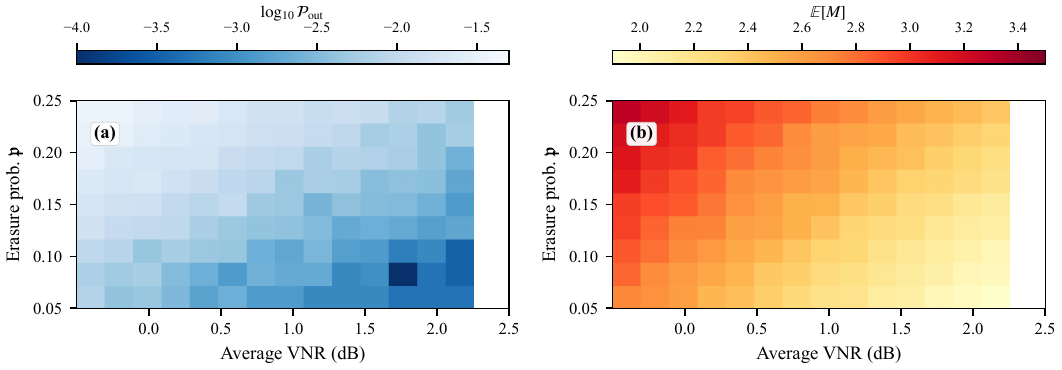}\label{fig:v2x_raptor_grid}}
\caption{V2X delivery under identical channel realizations ($M_{\max}=8$, $10^4$ trials at $\mathfrak{p}=0.15$; $6\times10^3$ trials per grid point): (a) block-Rayleigh outage and $\mathbb{E}[M]$ with fountain accumulation \eqref{eq:raptor_accum}; (b) Raptor-only $\mathcal{P}_{\mathrm{out}}$ and $\mathbb{E}[M]$ heatmaps.}
\end{figure*}

We evaluate the V2X channel model \eqref{eq:v2x_fading} under coherent detection after CSI estimation at legitimate receivers.
Each Monte Carlo trial emulates one broadcast session as in Theorem~\ref{thm:rateless_mono}: for slot $m=1,\ldots,M_{\max}$, an independent fading gain $|h_{\mathcal{D},m}|^2\sim\mathrm{Exp}(1)$ is drawn, the packet is erased with probability $\mathfrak{p}$, and otherwise a decode attempt is made at $\mathrm{VNR}_{\mathcal{D},m}$.
For matched-length LDLC and LDPC lattices, each attempt retransmits the same fixed block (H-ARQ diversity); success on attempt $m$ occurs with probability $1-\mathrm{FER}_{\mathrm{codec}}(\mathrm{VNR}_{\mathcal{D},m})$.
For the proposed Raptor lattice, each non-erased slot delivers a new fountain block; after $k$ received blocks the link-level decoder applies \eqref{eq:raptor_accum} to the Raptor AWGN $\mathrm{FER}(\mathrm{VNR})$ curve in Fig.~\ref{fig:101}.
The session succeeds once any attempt decodes; otherwise it is in outage with
\begin{equation}\label{eq:outage}
\begin{aligned}
\mathcal{P}_{\mathrm{out}}
&=\Pr\{\mathcal{S}_{M_{\max}}^c\}\\
&=\Pr\big\{\forall m\le M_{\max}:\,E_m^c\big\}.
\end{aligned}
\end{equation}
and decoding-completion overhead $\mathbb{E}[M]$.
We simulate $N_{\mathrm{trials}}=10^4$ sessions per $(\bar{\mathrm{VNR}},\mathfrak{p})$ with $M_{\max}=8$, sharing identical fading and erasure draws across codecs at each point.
Fig.~\ref{fig:v2x_fading} reports results at $\mathfrak{p}=0.15$; at $\bar{\mathrm{VNR}}=1.5$~dB, $\mathcal{P}_{\mathrm{out}}=0.3\%$ (Raptor), $1.2\%$ (LDLC), and $2.4\%$ (LDPC), with $\mathbb{E}[M]\approx 2.3$, $2.2$, and $2.5$ packets, respectively.
Fountain accumulation lets Raptor outperform both fixed-length baselines at this urban operating point despite comparable AWGN thresholds in Table~\ref{tab:lattice_parameters}.
Fig.~\ref{fig:v2x_raptor_grid} summarizes Raptor-only $\mathcal{P}_{\mathrm{out}}$ and $\mathbb{E}[M]$ over the $(\bar{\mathrm{VNR}},\mathfrak{p})$ operating region ($6\times10^3$ trials per grid point) as paired heatmaps with color scales. Table~\ref{tab:v2x_results} summarizes representative operating points from Fig.~\ref{fig:v2x_fading}; Eve failure probabilities correspond to Fig.~\ref{fig:v2x_pls} at $\mathfrak{p}=0.15$ with $\Delta\gamma=6$~dB plus $2$~dB DSSS loss.

\begin{table}[t]
    \centering
    \caption{Selected V2X Monte Carlo results ($M_{\max}=8$, $N_{\mathrm{trials}}=10^4$). Codec rows at $\mathfrak{p}=0.15$ share identical fading realizations; Raptor uses \eqref{eq:raptor_accum}.}
    \label{tab:v2x_results}
    \setlength{\tabcolsep}{3pt}
    \renewcommand{\arraystretch}{1.05}
    \resizebox{\linewidth}{!}{
    \begin{tabular}{@{}lccccc@{}}
        \toprule
        Codec & $\bar{\mathrm{VNR}}$ (dB) & $\mathfrak{p}$ & $\mathcal{P}_{\mathrm{out}}$ & $\mathbb{E}[M]$ & Eve fail. \\
        \midrule
        Raptor & 0.5 & 0.05 / 0.15 / 0.25 & 0.2\% / 0.8\% / 1.9\% & 2.40 / 2.60 / 2.89 & -- \\
        Raptor & 1.0 & 0.15 & 0.5\% & 2.47 & 70.8\% \\
        Raptor & 1.5 & 0.15 & 0.3\% & 2.30 & 62.2\% \\
        LDLC & 1.5 & 0.15 & 1.2\% & 2.22 & -- \\
        LDPC & 1.5 & 0.15 & 2.4\% & 2.52 & -- \\
        \bottomrule
    \end{tabular}
    }
    \vspace{-0.15cm}
\end{table}

\subsection{Secure Broadcast Evaluation}
\begin{figure}[tb]
    \centering
    \includegraphics[width=\columnwidth]{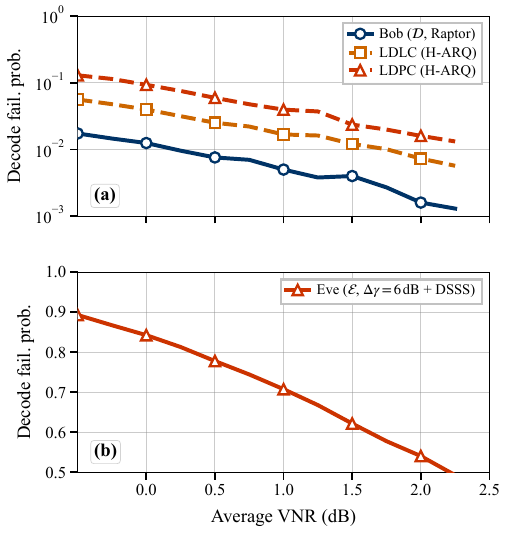}
    \caption{Decode failure within $M_{\max}=8$ packets ($\mathfrak{p}=0.15$, $\Delta\gamma=6$~dB plus $2$~dB DSSS loss at Eve): (a)~Bob (Raptor) with LDLC/LDPC H-ARQ baselines (log scale); (b)~Eve (Raptor, fountain accumulation \eqref{eq:raptor_accum} with $\eta=0.5$).}
    \label{fig:v2x_pls}
\end{figure}

The third evaluation tier measures session-level secure broadcast margin under Section~\ref{sec_v2x_protocol}: whether destination $\mathcal{D}$ completes fine-lattice decoding within $M_{\max}=8$ fountain packets while passive eavesdropper $\mathcal{E}$ remains far from success.
We identify Bob with $\mathcal{D}$ and Eve with $\mathcal{E}$ on the block-Rayleigh erasure channel of Section~\ref{sec_v2x}.
For each average-VNR point we draw $N_{\mathrm{trials}}=10^4$ paired sessions with identical sidelink fading and erasure realizations at both receivers; Eve's post-equalization SNR is reduced through \eqref{eq:gamma_E}.
Parameters are $\mathfrak{p}=0.15$, $\mathfrak{s}=64$, $\mathcal{L}=\lfloor\sqrt{(1-\mathfrak{p})\mathfrak{s}}\rfloor=7$, $\Delta\gamma=6$~dB, and $\Delta\gamma_{\mathrm{DSSS}}=2$~dB ($\Delta\gamma_{\mathrm{tot}}=8$~dB).
Since $(1-\mathfrak{p})\mathfrak{s}>1$, Proposition~\ref{prop:secrecy_rate} gives $R_s>0$; both receivers use AWGN $\mathrm{FER}(\mathrm{VNR})$ from Fig.~\ref{fig:101} with accumulation \eqref{eq:raptor_accum}, and Theorem~\ref{thm:secure_rateless}(iii) ensures $\Pr\{\mathcal{S}_{\mathcal{E},M_{\max}}\}\le \Pr\{\mathcal{S}_{\mathcal{D},M_{\max}}\}$.
Fig.~\ref{fig:v2x_pls} plots session decode failure $1-\Pr\{\mathcal{S}_{u,M_{\max}}\}$: panel~(a) shows Bob (Raptor) with LDLC/LDPC H-ARQ baselines from Fig.~\ref{fig:v2x_fading}; panel~(b) shows Eve with the $8$~dB penalty.
At $\bar{\mathrm{VNR}}=1.0$~dB, failures are $0.5\%$ (Bob) versus $70.8\%$ (Eve); at $1.5$~dB, $0.4\%$ versus $62.2\%$, confirming a large secure margin with near-reliable legitimate delivery.
Bob's $0.4\%$ differs slightly from Raptor outage $\mathcal{P}_{\mathrm{out}}=0.3\%$ in Fig.~\ref{fig:v2x_fading} because the sweeps use independent seeds; Eve fails less often than under a fixed-block model ($\approx96\%$ at $1.5$~dB) because she also accumulates fountain symbols via \eqref{eq:raptor_accum}, yet remains far from reliable decoding at the reduced SNR in \eqref{eq:gamma_E}.

\vspace{-0.12cm}

\subsection{Evaluation Scope and Limitations}\label{sec:eval_scope}
The numerical results support competitive AWGN performance relative to matched-length baselines, rateless delivery gains under block-Rayleigh fading and erasures, and a secure broadcast margin between $\mathcal{D}$ and $\mathcal{E}$.
Three modeling choices bound the conclusions.
The AWGN tier isolates physical-layer coding gain at matched block lengths.
In the fading tier, Raptor fountain accumulation \eqref{eq:raptor_accum} is contrasted with fixed-block H-ARQ for LDLC and LDPC lattices under identical channel realizations; full multistage belief-propagation decoding under time-varying CSI within a session is not reproduced.
The accumulation shift $\eta\,\Delta_{\mathrm{acc}}(k)$ is a conservative link-level surrogate calibrated at $\eta=0.5$ rather than a bit-exact fountain decoder.
Blockchain-guided relay selection is integrated architecturally in Section~\ref{sec_v2x_protocol} but not quantitatively benchmarked here.

\section{Conclusion}\label{sec_conclusion}
This paper presented a cooperative vehicle-to-everything broadcast framework that integrates rateless Construction-D$'$ Raptor lattices, nested-lattice physical-layer security with direct-sequence spread spectrum, and blockchain-assisted trust management for decode-and-forward relaying.
A block-Rayleigh sidelink model with random erasures and a passive eavesdropper was formulated, together with performance metrics tailored to vehicular broadcast, including outage probability, decoding-completion overhead, and secure broadcast margin.

The proposed multilevel lattice codec combines heuristic Luby-transform codes, quasi-cyclic low-density parity-check precodes, and girth-4 cycle removal within Construction-D$'$.
On an additive white Gaussian noise reference channel, the design achieves competitive symbol error rate and improved frame error rate relative to generalized Construction-D$'$ low-density parity-check and low-density lattice code baselines at comparable block lengths; at $n=995$, the FER gains are $0.07$~dB and $0.32$~dB over the matched-length LDLC and LDPC lattices, respectively.

Under block-Rayleigh fading and erasures, rateless-aware Monte Carlo evaluation with matched-length LDLC and LDPC baselines under identical channel realizations shows that fountain accumulation yields $\mathcal{P}_{\mathrm{out}}=0.3\%$ and $\mathbb{E}[M]\approx 2.3$ at $\bar{\mathrm{VNR}}=1.5$~dB and $\mathfrak{p}=0.15$, compared with $1.2\%$ and $2.4\%$ outage for LDLC and LDPC, respectively.
An eavesdropper subject to an $8$~dB combined channel-quality and spreading loss fails to decode in $62\%$ of sessions at the same operating point while legitimate delivery remains near-reliable.
These findings indicate that incremental fountain redundancy together with nested shaping and spread-spectrum hiding offers a favorable trade-off among sidelink occupancy, delivery reliability, and physical-layer confidentiality in urban V2X broadcast scenarios.
The numerical study validates the proposed codec under the rateless-aware link-level abstraction summarized in Section~\ref{sec:eval_scope}.
Promising directions for future work include NR-V2X link-level integration and trace-driven calibration of relay-trust statistics.

\bibliographystyle{IEEEtran}

\bibliography{Refer}

@ARTICLE{fan,
  author={Zhou, Fan and Kurkoski, Brian~M.},
  journal={IEEE Trans. on Commun.},
  title={Construction {D}’ Lattices for Power-Constrained Communications},
  year={2022},
  volume={70},
  number={4},
  pages={2200-2212},
  doi={10.1109/TCOMM.2022.3147235}}

@ARTICLE{silva,
  author={Branco {da Silva}, Paulo Ricardo and Silva, Danilo},
  journal={IEEE Trans. on Inf. Theory},
  title={Multilevel {LDPC} Lattices With Efficient Encoding and Decoding and a Generalization of {C}onstruction {$D^{\prime}$} },
  year={2019},
  volume={65},
  number={5},
  pages={3246-3260},
  doi={10.1109/TIT.2018.2883119}}

@ARTICLE{2019,
  author={Yu, Wei and Narayanan, Krishna and Cheng, Jun and Wu, Jun},
  journal={IEEE Commun. Letters},
  title={Raptor Codes With Descending Order Degrees for {AWGN} Channels},
  year={2020},
  volume={24},
  number={1},
  pages={29-33},
  doi={10.1109/LCOMM.2019.2948906}}

@ARTICLE{a,
  author={Khodaiemehr, Hassan and Sadeghi, Mohammad-Reza and Sakzad, Amin},
  journal={IEEE Trans. on Commun.},
  title={Practical Encoder and Decoder for Power Constrained {QC LDPC}-Lattice Codes},
  year={2017},
  volume={65},
  number={2},
  pages={486-500},
}

@book{11fan,
  author = {J.~H. Conway and N.~J.~A. Sloane},
  title = {Sphere Packing, Lattices and Groups},
  edition = {3rd},
  publisher = {Springer Verlag},
  address = {New York, NY, USA},
  year = {1999}
}

@ARTICLE{drshok,
  author={Shokrollahi, A.},
  journal={IEEE Trans. on Inf. Theory},
  title={Raptor codes},
  year={2006},
  volume={52},
  number={6},
  pages={2551-2567},
  doi={10.1109/TIT.2006.874390}}

@article{IET,
  author = {M. Zhang and S. Kim and Jin Chang and W. Kim},
  title = {Soft decoding method for systematic {Raptor} codes},
  journal = {IET Communications},
  month = {October},
  year = {2015}
}

@INPROCEEDINGS{13fan,
  author={Chen, Siyu and Kurkoski, Brian~M. and Rosnes, Eirik},
  booktitle={IEEE 10th International Symposium on Turbo Codes \& Iterative Information Processing (ISTC)},
  title={Construction ${D^{\prime}}$ Lattices from Quasi-Cyclic Low-Density Parity-Check Codes},
  year={2018},
  pages={1-5},
  doi={10.1109/ISTC.2018.8625376}}

@article{30fan,
  author = {R.~Y. Shao and S. Lin and M.~P.~C. Fossorier},
  title = {Two Decoding Algorithms for Tailing Codes},
  journal = {IEEE Trans. on Commun.},
  volume = {51},
  number = {10},
  pages = {1658--1665},
  month = {October},
  year = {2003}
}

@inproceedings{27fan,
  author = {F. Zhou and B.~M. Kurkoski},
  title = {Shaping Gain of Lattices Based on Convolutional Codes and {Construction A}},
  booktitle = {2018 International Symposium on Inf. Theory and Its Applications (ISITA)},
  location = {Singapore},
  year = {2018},
  pages = {183--187}
}

@book{28fan,
  author = {S.~I.~R. Costa and F. Oggier and A. Campello and J.~C. Belfiore and E. Viterbo},
  title = {Lattices Applied to Coding for Reliable and Secure Communications},
  series = {SpringerBriefs in Mathematics},
  publisher = {Springer},
  year = {2017},
  address = {Cham, Switzerland}
}

@ARTICLE{6fan,
  author={Erez, U. and Zamir, R.},
  journal={IEEE Trans. on Inf. Theory},
  title={Achieving $\frac{1}{2}\log (1+\mathrm{SNR})$ on the {AWGN} channel with lattice encoding and decoding},
  year={2004},
  volume={50},
  number={10},
  pages={2293-2314},
  doi={10.1109/TIT.2004.834787}}

@article{15fan,
  author = {B. Kudryashov and K. Yurkov},
  title = {Linear Code-Based Vector Quantization for Independent Random Variable},
  year = {2008},
  journal = {arXiv preprint},
  eprint = {arXiv:0805.2379},
  url = {https://arxiv.org/abs/0805.2379}
}

@INPROCEEDINGS{25fan,
  author={Vem, Avinash and Huang, Yu-Chih and Narayanan, Krishna~R. and Pfister, Henry~D.},
  booktitle={IEEE International Symposium on Inf. Theory},
  title={Multilevel lattices based on spatially-coupled {LDPC} codes with applications},
  year={2014},
  volume={},
  number={},
  pages={2336-2340},
  doi={10.1109/ISIT.2014.6875251}}

@ARTICLE{3kah,
  author={Poltyrev, G.},
  journal={IEEE Trans. on Inf. Theory},
  title={On coding without restrictions for the {AWGN} channel},
  year={1994},
  volume={40},
  number={2},
  pages={409-417},
  doi={10.1109/18.312163}}

@ARTICLE{4kah,
  author={Forney, G.~D. and Trott, M.~D. and Sae-Young Chung},
  journal={IEEE Trans. on Inf. Theory},
  title={Sphere-bound-achieving coset codes and multilevel coset codes},
  year={2000},
  volume={46},
  number={3},
  pages={820-850},
  doi={10.1109/18.841165}}

@ARTICLE{9kah,
  author={Khodaiemehr, Hassan and Kiani, Dariush and Sadeghi, Mohammad-Reza},
  journal={IEEE Trans. on Commun.},
  title={{LDPC} Lattice Codes for Full-Duplex Relay Channels},
  year={2017},
  volume={65},
  number={2},
  pages={536-548},
  }

@INPROCEEDINGS{10kah,
  author={Kurkoski, Brian~M. and Dauwels, Justin and Loeliger, Hans-Andrea},
  booktitle={IEEE International Symposium on Inf. Theory},
  title={Power-constrained communications using {LDLC} lattices},
  year={2009},
  volume={},
  number={},
  pages={739-743},
  doi={10.1109/ISIT.2009.5205635}}

@INPROCEEDINGS{13kah,
  author={di Pietro, Nicola and Boutros, Joseph~J. and Z\'{e}mor, Gilles and Brunel, Lo\"{i}c},
  booktitle={IEEE Inf. Theory Workshop},
  title={Integer low-density lattices based on {Construction A}},
  year={2012},
  volume={},
  number={},
  pages={422-426},
  doi={10.1109/ITW.2012.6404707}}

@ARTICLE{HassanIT,
  author={Khodaiemehr, Hassan and Panario, Daniel and Sadeghi, Mohammad-Reza},
  journal={IEEE Trans. on Inf. Theory},
  title={Design and Practical Decoding of Full-Diversity {C}onstruction {A} Lattices for Block-Fading Channels},
  year={2021},
  volume={67},
  number={1},
  pages={138-163},
  doi={10.1109/TIT.2020.3008973}}

@ARTICLE{20kah,
  author={Li, Z. and Chen, L. and Zeng, L. and Lin, S. and Fong, W.},
  journal={IEEE Trans. on Commun.},
  title={Efficient Encoding of Quasi-Cyclic Low-Density Parity-Check Codes},
  year={2005},
  volume={53},
  number={11},
  pages={1973-1973},
  doi={10.1109/TCOMM.2005.858628}}

@INPROCEEDINGS{14kah,
  author={Yanfei Yan and Cong Ling},
  booktitle={IEEE Inf. Theory Workshop},
  title={A Construction of lattices from polar codes},
  year={2012},
  volume={},
  number={},
  pages={124-128},
  doi={10.1109/ITW.2012.6404640}}

@ARTICLE{11kah,
  author={Sommer, Naftali and Feder, Meir and Shalvi, Ofir},
  journal={IEEE Trans. on Inf. Theory},
  title={Low-Density Lattice Codes},
  year={2008},
  volume={54},
  number={4},
  pages={1561-1585},
  doi={10.1109/TIT.2008.917684}}

@INPROCEEDINGS{18kah,
  author={{di Pietro}, Nicola and Basha, Nour and Boutros, Joseph J.},
  booktitle={IEEE Inf. Theory Workshop (ITW)},
  title={Non-binary {GLD} codes and their lattices},
  year={2015},
  volume={},
  number={},
  pages={1-5},
  doi={10.1109/ITW.2015.7133127}}

@inproceedings{19kah,
    author = {J.~J. Boutros and N. {di Pietro} and Y. C. Huang},
    title = {Spectral thinning in {GLD} lattice},
    booktitle = {Inform. Theory and Applications (ITA) Workshop},
    month = {January},
    year = {2015},
    pages = {1--9},
}

@ARTICLE{21kah,
  author={Nazer, Bobak and Gastpar, Michael},
  journal={IEEE Trans. on Inf. Theory},
  title={Compute-and-Forward: Harnessing Interference Through Structured Codes},
  year={2011},
  volume={57},
  number={10},
  pages={6463-6486},
  doi={10.1109/TIT.2011.2165816}}

@INPROCEEDINGS{4silva,
  author={Boutros, Joseph~J. and {di Pietro}, Nicola and Basha, Nour},
  booktitle={IEEE Inf. Theory Workshop (ITW)},
  title={Generalized low-density ({GLD}) lattices},
  year={2014},
  volume={},
  number={},
  pages={15-19},
  doi={10.1109/ITW.2014.6970783}}

@ARTICLE{19silva,
  author={Ling, Cong and Luzzi, Laura and Belfiore, Jean~Claude and Stehl\'{e}, Damien},
  journal={IEEE Trans. on Inf. Theory},
  title={Semantically Secure Lattice Codes for the {Gaussian} Wiretap Channel},
  year={2014},
  volume={60},
  number={10},
  pages={6399-6416},
  doi={10.1109/TIT.2014.2343226}}

@ARTICLE{21silva,
  author={Zhan, Jiening and Nazer, Bobak and Erez, Uri and Gastpar, Michael},
  journal={IEEE Trans. on Inf. Theory},
  title={Integer-Forcing Linear Receivers},
  year={2014},
  volume={60},
  number={12},
  pages={7661-7685},
  doi={10.1109/TIT.2014.2361782}}

@ARTICLE{18silva,
  author={Krithivasan, Dinesh and Pradhan, S.~Sandeep},
  journal={IEEE Trans. on Inf. Theory},
  title={Lattices for Distributed Source Coding: Jointly {Gaussian} Sources and Reconstruction of a Linear Function},
  year={2009},
  volume={55},
  number={12},
  pages={5628-5651},
  doi={10.1109/TIT.2009.2032853}}

@INPROCEEDINGS{18IET,
  author={Zhang, Meixiang and Kim, Sooyoung},
  booktitle={European Wireless 2014; 20th European Wireless Conference},
  title={A New Soft Decoding Method for Systematic {LT} Codes},
  year={2014},
  volume={},
  number={},
  pages={1-6},
  doi={}}

@INPROCEEDINGS{dretesami,
  author={Etesami, O. and Molkaraie, M. and Shokrollahi, A.},
  booktitle={International Symposium on Inf. Theory, 2004. ISIT 2004. Proceedings.},
  title={Raptor codes on symmetric channels},
  year={2004},
  volume={},
  number={},
  pages={38-},
  doi={10.1109/ISIT.2004.1365076}}

@misc{malema2025,
  author       = {Gabofetswe Malema},
  title        = {Flexible Girth 8 {QC-LDPC} Codes},
  year         = {2025},
  howpublished = {\url{https://www.mathworks.com/matlabcentral/fileexchange/24882-flexible-girth-8-qc-ldpc-codes}},
  note         = {MATLAB Central File Exchange},
  month        = {January},
  urldate      = {2025-01-27}
}

@ARTICLE{8my_thesis,
  author={Cai, Fang and Zhang, Xinmiao and Declercq, David and Planjery, Shiva~Kumar and Vasi\'{c}, Bane},
  journal={IEEE Trans. on Circuits and Systems I: Regular Papers},
  title={Finite Alphabet Iterative Decoders for {LDPC} Codes: Optimization, Architecture and Analysis},
  year={2014},
  volume={61},
  number={5},
  pages={1366-1375},
  doi={10.1109/TCSI.2014.2309896}}

@ARTICLE{sara,
  author={Jayasooriya, Sachini and Shirvanimoghaddam, Mahyar and Ong, Lawrence and Johnson, Sarah~J.},
  journal={IEEE Trans. on Commun.},
  title={Analysis and Design of {Raptor} Codes Using a Multi-Edge Framework},
  year={2017},
  volume={65},
  number={12},
  pages={5123-5136},
  doi={10.1109/TCOMM.2017.2750179}}

@ARTICLE{29silva,
  author={Parrao Hernandez, Ricardo Antonio and Kurkoski, Brian~M.},
  journal={IEEE Trans. on Commun.},
  title={The Three/Two {Gaussian} Parametric {LDLC} Lattice Decoding Algorithm and Its Analysis},
  year={2016},
  volume={64},
  number={9},
  pages={3624-3633},
  doi={10.1109/TCOMM.2016.2594286}}

@INPROCEEDINGS{CCECE,
  author={Sharifi, Pegah and Bagheri, Khadijeh and Khodaiemehr, Hassan and Feng, Chen},
  booktitle={Canadian Conference on Electrical and Computer Engineering and Industry Summit (CCECE)},
  title={Construction-{D}$'$ Lattices Based on Raptor Codes for Unconstrained {AWGN} Channels},
  year={2025},
  pages={522--526},
   doi={10.1109/CCECE64018.2025.11364450},
}

@article{ghourab2023,
  author    = {Esraa~M. Ghourab and Wael Jaafar and Lina Bariah and Sami Muhaidat and Halim Yanikomeroglu and Shimaa Naser},
  title     = {Interplay between Physical Layer Security and Blockchain Technology for {5G} and Beyond: A Comprehensive Survey},
  journal   = {TechRxiv},
  year      = {2023},
  month     = {March},
  doi       = {10.36227/techrxiv.21601848.v2},
  note      = {Preprint}
}

@article{ref202,
  author = {E.~M. Ghourab and M. Azab and N. Ezzeldin},
  title = {Blockchain-guided dynamic best-relay selection for trustworthy vehicular communications},
  journal = {IEEE Trans. on Intelligent Transportation Systems},
  volume = {28},
  number = {8},
  pages = {13678--13693},
  month = {August},
  year = {2022}
}

@ARTICLE{ref203,
  author={Ghourab, Esraa~M. and Bariah, Lina and Muhaidat, Sami and Sofotasios, Paschalis~C. and {Al-Qutayri}, Mahmoud and Damiani, Ernesto},
  journal={IEEE Open Journal of Vehicular Technology},
  title={Reputation-Aware Relay Selection With Opportunistic Spectrum Access: A Blockchain Approach},
  year={2023},
  volume={4},
  pages={389-403},
  doi={10.1109/OJVT.2023.3263804}}

@ARTICLE{Erasure,
  author={Rahman, Saifur and Feng, Chen and Chaaban, Anas},
  journal={IEEE Commun. Letters},
  title={A Simplified Approach to Achieving the Capacity of the {AWGN} Channel With Erasures Using Nested Lattice Codes},
  year={2022},
  volume={26},
  number={3},
  pages={509-512},
  doi={10.1109/LCOMM.2021.3135867}}

@ARTICLE{Verdu,
  author={Verdu, Sergio and Weissman, Tsachy},
  journal={IEEE Trans. on Inf. Theory},
  title={The Information Lost in Erasures},
  year={2008},
  volume={54},
  number={11},
  pages={5030-5058},
  doi={10.1109/TIT.2008.929968}}

@book{golomb1967shift,
  title={Shift Register Sequences},
  author={Golomb, S.~W.},
  lccn={66017667},
  series={Holden-Day series in information systems},
  url={https://books.google.ca/books?id=LqtMAAAAMAAJ},
  year={1967},
  publisher={Holden-Day}
}

@book{DSSS,
  author    = {Peterson, R.~L. and Ziemer, R.~E. and Borth, D.~E.},
  title     = {Introduction to Spread Spectrum Communications},
  year      = {1995},
  publisher = {Prentice-Hall},
  address   = {Englewood Cliffs, NJ}
}

@misc{raptorcast2025,
  author       = {{Category Labs}},
  title        = {RaptorCast: Designing a Messaging Layer},
  year         = {2025},
  url          ={https://www.category.xyz/blogs/raptorcast-designing-a-messaging-layer},
  note         = {{Accessed}: 2025-10-25}
}

@ARTICLE{Polarm,
  author={Liu, Ling and Yan, Yanfei and Ling, Cong and Wu, Xiaofu},
  journal={IEEE Trans. on Commun.},
  title={Construction of Capacity-Achieving Lattice Codes: Polar Lattices},
  year={2019},
  volume={67},
  number={2},
  pages={915-928},
  doi={10.1109/TCOMM.2018.2876113}}

@article{Pietro2018,
  author    = {N. {di Pietro} and G. Z\'{e}mor and J.~J. Boutros},
  title     = {{LDA} lattices without dithering achieve capacity on the {Gaussian} channel},
  journal   = {IEEE Trans. on Inf. Theory},
  volume    = {64},
  number    = {3},
  pages     = {1561--1594},
  month     = {March},
  year      = {2018},
}

@ARTICLE{Wyner,
  author={Wyner, A.~D.},
  journal={The Bell System Technical Journal},
  title={The wire-tap channel},
  year={1975},
  volume={54},
  number={8},
  pages={1355-1387},
  doi={10.1002/j.1538-7305.1975.tb02040.x},}

@ARTICLE{Maryam,
  author={Sadeghi, Maryam and Khodaiemehr, Hassan and Feng, Chen},
  journal={IEEE Trans. on Commun.},
  title={Full-Diversity {Construction-D} Lattices: Design and Decoding Perspective on Block-Fading Channels},
  year={2025},
  volume={73},
  number={12},
  pages={13146-13162},
  doi={10.1109/TCOMM.2025.3615774},}

\end{document}